\documentclass[aps,preprintnumbers,twocolumn,showpacs]{revtex4}
\usepackage{amsmath}
\usepackage{mathtools}
\usepackage{amssymb}
\usepackage{epsfig,epsf}
\usepackage{graphicx}
\usepackage{color,pstricks}
\usepackage{wasysym}

\begin{document}

\title{Dirac Green's function approach to graphene-superconductor junctions with well defined edges}
\author{William J. Herrera$^1$, P. Burset$^2$, and A. Levy Yeyati$^2$}
\affiliation{$^1$Departamento de F\'{\i}sica, Universidad Nacional de Colombia, Bogot\'{a},
Colombia\\
$^2$Departamento de F\'{\i}sica Te\'{o}rica de la Materia Condensada C-V,
Facultad de Ciencias, Universidad Aut\'{o}noma de Madrid, E-28049 Madrid,
Spain } 
\date{\today }

\begin{abstract}
This work presents a novel approach to describe spectral properties of graphene layers with well defined edges. We microscopically analyze the boundary problem for the continuous Bogoliubov-de Gennes-Dirac (BdGD) equations and derive the Green functions for normal and superconducting graphene layers. Importing the idea used in tight-binding (TB) models of a microscopic hopping that couples different regions, we are able to set up and solve an algebraic Dyson's equation describing a graphene-superconductor junction. For this coupled system we analytically derive the Green functions and use them to calculate the local density of states and the spatial variation of the induced pairing correlations in the normal region. Signatures of specular Andreev reflections are identified.
\end{abstract}

\maketitle

\section{Introduction}

Ideally, an interface between a graphene layer and a superconductor provides
a unique opportunity to study the interplay between a relativistic-like
band-structure and pairing interactions between electrons and holes. The
advances in the techniques of isolation and manipulation of graphene layers
together with the possibility to deposit nanoscale superconducting
electrodes on top of them is bringing this idealized situation closer \cite{heersche07} .

Several theoretical studies of the electronic and transport properties of
graphene-superconductor junctions have been reported in recent years. As in
the case of normal metals, a key ingredient for understanding these
properties is the concept of Andreev reflection \cite{Andreev} by means of which an incident electron on the interface is reflected as a hole while a Cooper
pair is created on the superconductor. In graphene, however, Andreev
reflections may couple electrons in the conduction band with holes in the
valence band, which leads to a \textit{specular} reflection, as opposed to
the usual \textit{retro} Andreev reflection \cite{beenakker06}. Very recently we have shown that interface bound states (IBSs) appear for energies within the superconducting gap due to the interplay of Andreev and normal reflection in
graphene-superconductor junctions \cite{burset09}.

Within these studies various different methods and approximations have been
applied. In particular, the Bogoliubov-de Gennes-Dirac equations (BdGD) have been solved under the approximation of neglecting intervalley
scattering and applied to the calculation of the differential conductance and spectral properties in graphene-superconductor (GS) junctions \cite{beenakker06,GS}. Other works have extended this
method to situations including an insulator (GIS junctions \cite{GIS}) or to the study of the Josephson effect in SGS junctions \cite{SGS}. In spite of their interest these works are limited in the sense that they cannot provide an answer on the effect of the atomic scale properties of the interfaces or what would happen in the presence of well defined edges in the graphene layers. This is an
important issue because it is well known that the electronic spectrum of a
graphene nanoribbon is strongly dependent on the direction in which its
edges are defined. Armchair and zig-zag cases have been the most studied
but one can also envisage and study edges along arbitrary directions. From
the point of view of experiments it is also becoming possible to study
transport properties in graphene nanoribbons with well defined edges at the
atomic scale \cite{exp-edges}.

These type of studies typically require the use of microscopic
tight-binding (TB) models and some of them have been conducted in the
case of GS junctions \cite{burset08,TB-num}. Although such methods are usually implemented numerically, much progress has been made in the derivation of analytical expressions within TB models combined with Green function techniques \cite{burset08}.
However, these expressions are typically more complex than those which would be obtained within the BdGD model.

The aim of the present work is to go one step beyond the existing works \cite{tkachov09} and develop a method based on the BdGD model which could deal with the presence of well defined edges or interfaces at the atomic scale. Our method starts from the determination of the Green functions for the BdGD equations in the case of isolated graphene or superconducting layers. In order to fix the boundary conditions for these isolated regions it is in general necessary to mix solutions corresponding to different valleys. For the purpose of illustrating the method we would consider separately the cases of armchair
and zig-zag edges. For obtaining the properties of a coupled
graphene-graphene or graphene-superconductor region we define a short range
coupling potential which mimics the usual hopping term in the TB
models and then solve the corresponding Dyson equation analytically for
several junctions. As a test we show how known results are recovered in
different limits. We further use the method to study the spatial properties of the local density of states (LDOS) and the induced pairing correlations on a graphene layer. We show that, on top of an atomic scale modulation, these spatial correlations exhibit signatures of specular Andreev reflections. They also display a long range decay due to the presence of the above mentioned IBS.

This work is organized as follows: in the next section we implement Green's functions for graphene nanoribbons with armchair or zigzag edges. In section III it is explained how we model the microscopic link between layers and solve the corresponding Dyson's equation. Finally, section IV covers the coupling between a normal sheet of graphene and a superconducting one. Analytical expressions for the Green functions of the coupled system are given. Profiles for the LDOS and the spatial behavior of the induced superconducting pairing correlations are presented as well. This paper is closed by some concluding remarks. Furthermore, an appendix has been added to show details of the calculations.

\section{Green's function approach for graphene layers with edges}

\begin{figure}
\includegraphics[width=8.5cm]{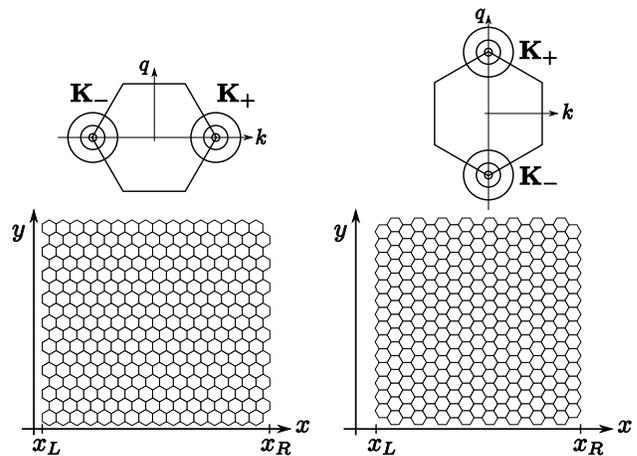}
\caption{Geometry of the systems treated in this work: a horizontally finite sheet of graphene with armchair edges (left) or zigzag edges (right). The sheets are assumed to be infinite in the vertical direction. The edges are located at positions $x_L$ and $x_R$. These positions can be taken to be infinite, representing a bulk of graphene (infinite plane) or a semi-infinite layer (infinite half-plane). For each geometry we represent the chosen Dirac points in the Brillouin zone.}
\label{fig:bulk_edges}
\end{figure}

Graphene is a single layer of carbon atoms arranged in a honeycomb structure. Its hexagonal lattice is constructed as the superposition of two triangular lattices $A$ and $B$. A nearest neighbor TB model for graphene can be, in the continuum limit, represented by a linearized Hamiltonian with six Dirac points at each corner of the hexagonal Brillouin zone. Only two of these points are inequivalent which, for the orientation of the layer shown in the left panel of Fig. \ref{fig:bulk_edges}, can be assumed to lie in the $x$-axis and denoted as $\mathbf{K}_{\pm }=\left(\pm K,0\right)$. Single-particle excitations around these points are represented by a Dirac Hamiltonian like $\hat{H}_{\pm }=\hbar v \mathbf{\sigma}_{\pm}  \cdot \mathbf{k} -E_{F}\hat{\sigma}_{0} = \hbar v\left[ \pm k\hat{\sigma}_{x}+q\hat{\sigma}_{y}\right] -E_{F}\hat{\sigma}_{0}$, where $\hat{\sigma}_{x,y}$ are the Pauli matrices (with $\sigma _{0}$ the $2\times 2$ identity matrix) acting on graphene's lattice subspace, $E_{F}$ is the Fermi energy and the velocity $v=\sqrt{3}t_{g}a/2\hbar \approx 10^{6}$ m/s is proportional to the lattice constant $a=0.246$ nm and to the nearest-neighbor hopping energy $t_{g}\approx 3$ eV on the honeycomb lattice of carbon atoms. With this choice $K = 4\pi /3a$. The degree of freedom due to the $A-B$ lattices define a \textit{pseudospin} which points in the direction of motion of the particle and the pseudospin operator $\mathbf{\sigma}_{\pm} = \left(\pm \hat{\sigma}_{x}, \hat{\sigma}_{y} \right)$ is used to define the Hamiltonian on each valley. Finally, $\mathbf{k}$ is the wave vector measured from each Dirac point $\mathbf{K}_{\pm }$.

Wave functions around each Dirac point satisfy the equation $\hat{H}_{\pm} \Phi_{\pm}\left( x,y\right) = E \Phi_{\pm}\left( x,y\right)$, with $E>0$ being the excitation energy of an electron-like quasiparticle. In the present work we consider clean samples which are infinite along the $y$-direction but may have edges or interfaces on $x$-direction (see Fig. \ref{fig:bulk_edges}). Thus, momentum along the $y$-axis ($\hbar q$) is conserved and the wave function can be written as $\Phi_{\pm}(x,y)=e^{iqy}\phi_{\pm} (x)$. With the replacement $\mathbf{k}\left( q\right) =\left(-i \partial _{x}, q\right)$ in Dirac's equation, linear independent solutions for each valley are 
\begin{eqnarray}
\phi_{+}(x) &=& c^{+}_1 e^{ikx}\varphi _{1}+ c^{+}_2 e^{-ikx}\varphi _{2} \\
\phi_{-}(x) &=& c^{-}_2 e^{ikx}\varphi_{2} + c^{-}_1 e^{-ikx}\varphi_{1} ,
\end{eqnarray}
with
\begin{equation}
\begin{array}{cc}
\varphi_{1} = \left( 
\begin{array}{c}
e^{-i\alpha /2} \\ 
e^{i\alpha /2}
\end{array}
\right) ,
&
\varphi_{2} = \left( 
\begin{array}{c}
e^{i\alpha /2} \\ 
-e^{-i\alpha /2}
\end{array}
\right) 
\end{array}
\label{eq:basis}
\end{equation}
and $e^{\pm i\alpha }=\hbar v (k\pm iq)/(E_{F}+E)$. The constants $c^{\pm}_{1,2}$ are determined by boundary conditions for the wave functions $\phi_{\pm} \left( x\right)$.

A complete description of the low energy electron excitations on a graphene layer is reached by superposing solutions on both valleys. Each solution $\Phi_{\pm}$ is modulated by a rapidly varying plane wave from each valley $e^{i\mathbf{K}_{\pm}\cdot \mathbf{r}}$ \cite{ando05} thus giving
\begin{equation}
\Phi \left(x,y \right)\! =\! e^{iqy}\phi (x)\! =\! e^{iqy}\! \left[ e^{iKx} \phi_{+}\!\left(x\right)\! +\! e^{-iKx} \phi_{-}\!\left(x \right) \right]\!.
\label{eq:fdo_superposition}
\end{equation}
This wave function is solution of Schr\"{o}dinger's equation for a Hamiltonian that combines both valleys. However, it is helpful to use a four-component spinor notation for the wave function made from the two-dimensional spinors for each valley in which the phase factors are omitted (i.e. $\psi^{T}=\left(\phi_{+}^{T},\phi_{-}^{T} \right)=\left(\phi_{+}^{A},\phi_{+}^{B},\phi_{-}^{A},\phi_{-}^{B}\right)$) \cite{TBQED}. This new notation is useful in order to calculate the full Green function of the system, including both valley and pseudospin degrees of freedom. The four-dimensional spinor obeys the equation $\check{H}\psi \left( x\right) =E\psi \left( x\right)$, where 
\begin{equation}
\check{H}=\left( 
\begin{array}{cc}
\hat{H}_{+} & 0 \\ 
0 & \hat{H}_{-}
\end{array}
\right)
\label{eq:hamiltonian}
\end{equation}
is the Dirac Hamiltonian in sublattice and valley spaces. We are using the notation $\check{\cdots}$ for $4\times 4$ matrices and $\hat{\cdots}$ for $2\times 2$ matrices. 

The Green function associated to $\Psi(x,y)=e^{iqy}\psi(x)$ is $\check{G}_{\psi}\left( x,x^{\prime },y\right) =\int \check{G}_{\psi}\left(x,x^{\prime };q\right) e^{iqy}dq$, where $\check{G}_{\psi}\left(x,x^{\prime };q\right) $ satisfies the $4 \times 4$ matrix equation
\begin{equation}
\begin{array}{c}
\left[\check{H}\left(q\right)-E\check{I} \right]  \check{G}_{\psi}\left( x,x^{\prime };q\right) = \delta (x-x^{\prime })  \check{I}   \\
\left(\! \begin{array}{cc} 
\hat{H}_{+}\!-\!E & 0 \\ 0 & \hat{H}_{-}\!-\!E 
\end{array} \! \right)
\left( \! \begin{array}{cc}
\hat{G}_{\psi}^{++} & \hat{G}_{\psi}^{+-} \\ 
\hat{G}_{\psi}^{-+} &\hat{G}_{\psi}^{--} 
\end{array} \! \right)\! =\! \delta (x-x^{\prime }) \check{I} , 
\end{array}
\label{eq:Green_dirac}
\end{equation}
with $\check{I}$ the 4-dimensional identity matrix. Hereafter we implicitly assume that $E$ stands for $E+i\eta$ and that the limit $\eta \rightarrow 0$ has been taken to obtain the retarded component of the Green function. From the elements of the full Green function $\check{G}_{\psi}$ we can define a valley superposed Green function $\hat{G}_{\phi}$, associated with the wave function of Eq. (\ref{eq:fdo_superposition}), as
\begin{eqnarray}
\hat{G}_{\phi}\left(x,x',y,y'\right) = \int \mathrm{d}q e^{iq \left(y-y'\right)} \times \nonumber \\
\sum\limits_{\mu , \nu =+,-} e^{i \left( \mathbf{K}_{\mu} \cdot \mathbf{r} - \mathbf{K}_{\nu} \cdot \mathbf{r}' \right) }  \hat{G}^{\mu \nu}_{\psi} \left(x,x';q\right) .
\label{eq:real_FdG}
\end{eqnarray}
Besides allowing valley mixing, this Green function include the phase factors $\exp{i \left( \mathbf{K}_{\mu} \cdot \mathbf{r} - \mathbf{K}_{\nu} \cdot \mathbf{r}' \right) }$ which are crucial to describe the presence of well defined edges or interfaces at the atomic scale. This valley superposed Green function is derived in the next sections for armchair and zigzag edges.

\begin{figure}
\includegraphics[width=8.5cm]{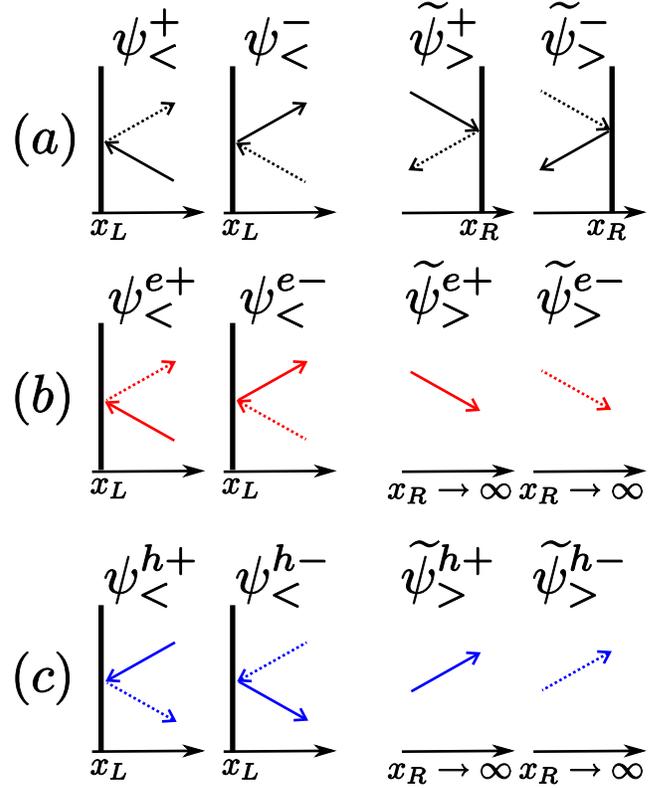}
\caption{(Color online) Schematic representation of the asymptotic solutions of Eq. (\ref{eq:Green_dirac}) for a system with armchair edges. Scattering at an armchair edge always changes the valley projection. The processes with non zero probability combine solutions on the left edge with one valley projection (e.g. $\psi_{<}^{+}$) with solutions on the right with the opposite valley index ($\widetilde{\psi}_{>}^{-}$). (a) Example of valley mixing at armchair edges ($x_L$ and $x_R$) for incident particles with a fixed valley index, where solid (dashed) lines represent particles with valley index $+$ ($-$). Reflection processes for electron (b) and hole-like (c) quasiparticles. For a semi-infinite region, $x_R$ goes to infinity and the asymptotic solutions are outgoing plane waves. For zigzag edges we have the same behavior without the change in the valley index.}
\label{fig:solutions_ribbon}
\end{figure}

The method that we implement to calculate the Green function is based in using the asymptotic solutions of the differential equation (\ref{eq:Green_dirac}). Basically, we extend the method used in Ref. \cite{tanaka} to the relativistic Bogoliubov-de Gennes equations. Thus, the full Green function in pseudospin and valley spaces can be written as 
\begin{equation}
\begin{array}{c}
\check{G}_{\psi}\left( x,x^{\prime };q\right) = \\
\left\{ 
\begin{array}{cr}
\sum\limits_{\mu ,\nu =+,-}A_{\mu \nu }\psi_{<}^{\mu }\left( x\right) \cdot  \widetilde{\psi}_{>}^{\nu T} \left( x^{\prime}\right) \cdot \check{\gamma} & ; x<x^{\prime} \\ 
\sum\limits_{\mu ,\nu = +,- }A_{\mu \nu }^{\prime} \psi_{>}^{\mu}\left( x\right) \cdot \widetilde{\psi}_{<}^{\nu T} \left( x^{\prime}\right) \cdot \check{\gamma} &; x>x^{\prime}
\end{array} \right. ,
\end{array}
\label{eq:FdG_full}
\end{equation}
where the indexes $\mu$ and $\nu$ represent different valley projections of the asymptotic solutions. We build the Green's function using tensor products as $\hat{G}(x \lessgtr
x^{\prime}) \propto \psi_{<(>)}^{\mu}(x) \cdot \widetilde{\psi}_{>(<)}^{\nu T}(x^{\prime})$. Where $\psi^{\mu}_{<}(x)$ and $\psi^{\nu}_{>}(x)$ are the asymptotic solutions that fulfill the boundary conditions to the left and right edges of the system, respectively. In Fig. \ref{fig:solutions_ribbon} $(a)$ the possible processes for an armchair edge are shown. The labels $\mu$ and $\nu$ are always determined by the valley index of the incident particle. On the other hand, $\widetilde{\psi}_{<(>)}^{\nu}(x')$ represent asymptotic solutions for the transposed Dirac's equation with the same boundary conditions and opposite wave number $\widetilde{\psi}_{<(>)}^{\nu} (\mathbf{r}) = \exp (-iqy)\widetilde{\psi}_{<(>)}^{\nu}(x)$. Since $\check{H}^{T}\left( -q\right) =\check{H}\left( q\right) $, the wave functions $\widetilde{\psi }_{<(>)}^{\nu T}(x)$ fulfill the same differential equation than $\psi^{\nu}_{<(>)}(x)$. The constants $A_{\mu \nu }^{(\prime )}$ are determined from the condition 
\begin{equation}
\begin{array}{c}
\sum\limits_{\mu , \nu = +,-}A_{\mu \nu }^{\prime }\psi_{>}^{\mu }\left( x\right) \cdot \psi_{<}^{\nu T}\left( x\right) \cdot \check{\gamma}  \\ -\sum\limits_{\mu , \nu = +,- }A_{\mu \nu } \psi_{<}^{\mu }\left( x\right) \cdot \psi_{>}^{\nu T}\left( x\right) \cdot \check{\gamma}=-\frac{i}{\hbar v} \hat{\tau}_{z} \otimes \hat{\sigma}_{x} ,
\label{eq:BC_AB_Dirac}
\end{array}
\end{equation}
which is obtained integrating Eq. (\ref{eq:Green_dirac}) around $x=x'$.

In our definition of the Green functions in Eq. (\ref{eq:FdG_full}) we have introduced the parity matrix $\check{\gamma}$ in order to make the scalar
product $\bar{\psi}\hat{G}\psi $ invariant under Lorentz transformations and
spatial inversion (parity transformation). Here $\bar{\psi}$ is the \textit{adjoint} Dirac spinor defined as $\bar{\psi}=\psi ^{\dag }\check{\gamma}$. The transformation rule for Dirac spinors is $\psi \rightarrow \psi ^{\prime }=\check{S}\psi $, where $\check{S}$ is a $4\times 4$ matrix that fulfills $\check{S}^{\dagger }\check{\gamma}\check{S}=\check{\gamma}$. With this rule, the
product $\left( \bar{\psi}\psi \right) ^{\prime }=\bar{\psi}\check{S}^{\dagger }\check{\gamma}\check{S}\psi ^{\prime }=\bar{\psi}\psi $ is Lorentz invariant (for a complete description of the symmetries of Dirac fermions see Ref. \cite{TBQED}).
Now, since $\check{G}\propto \psi \psi ^{T}\check{\gamma}$, we have that $\check{G}\rightarrow \check{G}^{\prime }=\check{S}\check{G}\check{\gamma}\check{S}\check{\gamma}$ and the product $\left( \bar{\psi}\check{G}\psi
\right) ^{\prime }=\psi ^{\dag }\check{S}\check{\gamma}\left( \check{S}\check{G}\check{\gamma}\check{S}\check{\gamma}\right) \check{\gamma}\psi =\psi ^{\dag }\check{\gamma}\check{G}\psi =\bar{\psi}\check{G}\psi $ is also Lorentz invariant. The matrix $\check{\gamma}$ in reciprocal space changes both valley and pseudospin indexes. Then, since the spinor is in the basis $(A_+,B_+,A_-,B_-)$ the matrix $\check{\gamma}$ can be written as $\check{\gamma}=\hat{\tau}_{x}\otimes \hat{\sigma}_{x}$ and we can check that the
Hamiltonian $\check{H}$ is transformed with $\check{\gamma}$ as $\check{\gamma}\check{H}\left( \mathbf{p}\right) \check{\gamma}=\check{H}\left( -\mathbf{p}\right) $. In the next section, where the armchair edge is studied, we use this form of the matrix $\check{\gamma}$. A simplified form of matrix $\check{\gamma}$ for one valley is derived in Appendix \ref{appendix_A} and is used for zigzag edges. 

\subsection{Armchair edges}

We can now apply this method to derive the Green functions for a layer of
graphene with armchair edges. Within the geometry depicted in the left panel of Fig. \ref{fig:bulk_edges}, the direction perpendicular to the edge may be infinite or have a finite size $W$. In this later case, we have two edges at positions $x_{L}$ and $x_{R}$. The main characteristic of an armchair edge is that scattering at the edges changes the valley index. This is a direct result of vanishing either the wave function (Eq. (\ref{eq:fdo_superposition})) or its derivative on both lattices at the armchair edge. 
Asymptotic solutions for this system correspond to incoming waves from one valley that are reflected on the other valley. All possible processes are illustrated in Fig. \ref{fig:solutions_ribbon}$(a)$. The index $\pm$ symbolizes an incoming wave around the Dirac point $\mathbf{K}_{\pm}$ that is reflected at the interface. For incident waves reflected at the left ($<$) or at the right ($>$) edge we have the four-dimensional spinors in valley and lattice spaces 
\begin{eqnarray}
\psi^{+}_{<(>)}(x) &=& \nonumber \\ e^{\mp ikx} \left( 
\begin{array}{c}
\varphi _{2(1)} \\ 
0
\end{array}
\right) &+& r_{L(R)}^{+} e^{\pm ikx}\left( 
\begin{array}{c}
0 \\ 
\varphi _{2(1)}
\end{array}
\right) \\
\psi^{-}_{<(>)}(x) &=& \nonumber \\  e^{\mp ikx}\left( 
\begin{array}{c}
0 \\ 
\varphi _{1(2)}
\end{array}
\right) &+& r_{L(R)}^{-} e^{\pm ikx}\left( 
\begin{array}{c}
\varphi _{1(2)} \\ 
0
\end{array}
\right)
\end{eqnarray}

The reflection amplitudes $r_{L(R)}^{\pm}$ are calculated imposing
boundary conditions to the two-dimensional wave function of Eq. (\ref{eq:fdo_superposition}) at the edge $x_{L(R)}$, this is $\phi(x_L)=\phi(x_R)=0$ or $\partial_x\phi(x_L)=\partial_x\phi(x_R)=0$, which yields
\begin{eqnarray}
r_{R}^{\pm} &=& s e^{\pm 2i\left( K\mp k\right) x_{R}} \nonumber \\
r_{L}^{\pm} &=& s e^{\pm 2i\left( K\pm k\right) x_{L}} ,
\label{eq:ref_coefs}
\end{eqnarray}
where $s=\pm $ corresponds to the case in which the wave function ($-$) or its derivative ($+$) is vanished at the interface. The main differences
between both cases are studied in the next section. For simplicity we
have chosen that $x_{L}\in (-\infty ,0]$ and $x_{R}\in \lbrack 0,\infty )$.
Combining the asymptotic solutions, as has been explained in the previous section, we obtain a general expression for the full Green function with armchair edges ($\check{G}_{\psi}^{\text{arm}}$) equivalent to Eq. (\ref{eq:FdG_full}).
The constants $A_{\mu =\nu}^{(\prime )}$ in this equation correspond to processes in which the excitation changes its valley projection when it
propagates from one edge to the other. Therefore, these processes have zero probability and, from Eq. (\ref{eq:BC_AB_Dirac}), one explicitly obtains $A_{++}^{(\prime )}=A_{--}^{(\prime )}=0$, $A_{+-}=A_{+-}^{\prime }=-i/\left( 2\hbar v\cos{\alpha} \left( 1-r_{L}^{+} r_{R}^{-}\right)
\right)$ and $A_{-+}=A_{-+}^{\prime }=-i/\left( 2\hbar v\cos{\alpha} \left(1-r_{L}^{-}r_{R}^{+}\right) \right)$. Substituting these results in $\check{G}_{\psi}^{\text{arm}}$ we can derive the valley superposed Green function for a layer of graphene with armchair edges
\begin{eqnarray}
&&\!\!\!\hat{G}^{\text{arm}}_{\phi}\left( x,x^{\prime }\right) = \frac{-i}{2\hbar v\cos{\alpha} \left(1-r_{L}^{-}r_{R}^{+}\right) } \left[ e^{i\left( K+k\right) \left\vert x-x^{\prime}\right\vert } \right. \nonumber \\
&&\left. + r_{L}^{-} r_{R}^{+} e^{-i\left( K+k\right)\left\vert x-x^{\prime}\right\vert } + r_{L}^{-} e^{i\left( K+k\right) (x+x^{\prime})} \right. \nonumber \\
&&\left. + r_{R}^{+} e^{-i\left( K+k\right) (x+x^{\prime })} \right] \varphi _{1}\varphi _{1}^{\dagger } + \frac{-i}{2\hbar v\cos{\alpha} \left( 1-r_{L}^{+} r_{R}^{-}\right) } \nonumber \\
&&\times \left[e^{-i\left( K-k\right) \left\vert x-x^{\prime}\right\vert} + r_{L}^{+}r_{R}^{-} e^{i\left( K-k\right) \left\vert x-x^{\prime}\right\vert } \right. \nonumber \\
&&\left. + r_{L}^{+} e^{-i\left( K-k\right) (x+x^{\prime })}+ r_{R}^{-} e^{i\left(
K-k\right) (x+x^{\prime })} \right] \varphi _{2}\varphi _{2}^{\dagger } .
\label{eq:FdG_arm_f}
\end{eqnarray}
The poles of the Green function are determined by $\cos {\alpha }=0$, $\left(1 - r_{L}^{-} r_{R}^{+}\right) = 0$ and $\left( 1-r_{L}^{+} r_{R}^{-}\right) =0$. The first one results in the bulk dispersion relation $\hbar
v\left\vert q\right\vert =\left\vert E_{F}+E\right\vert$. In the second and
third cases we reach the condition $2(K_{\pm }+k_{n})W=2n\pi $, with $W=x_{R}-x_{L}$ and $n$ an integer. Thus, the allowed values of the transverse momentum for a finite armchair sheet are $k_{n}=\frac{n\pi }{W}\mp \frac{4\pi }{3a}$, independently of the boundary condition chosen for the armchair edges. This result is in agreement with previous works on graphene nanoribbons with armchair edges \cite{brey06}.

We now consider the case of a superconducting graphene region, where the description of electron and hole excitations is done within Nambu space. We assume a s-wave pairing which leads to a constant gap $\Delta$ which is diagonal in sublattice space. The system is then represented by the BdGD equation written in lattice, Nambu and valley spaces ($8\times 8$ matrix). It reads 
\begin{eqnarray}
&&\left( 
\begin{array}{cccc}
\hat{H}_{+}-E_{F}^{S} & \Delta & 0 & 0 \\ 
\Delta^{*} & E_{F}^{S}-\hat{H}_{+} & 0 & 0 \\ 
0 & 0 & \hat{H}_{-}-E_{F}^{S} & \Delta \\ 
0 & 0 & \Delta^{*} & E_{F}^{S}-\hat{H}_{-}
\end{array}
\right) \nonumber \\
&&\;\;\;\;\times\left( 
\begin{array}{c}
\phi^{e}_{+} \\ 
\phi^{h}_{-} \\ 
\phi^{e}_{-} \\ 
\phi^{h}_{+}
\end{array}
\right) =E\left( 
\begin{array}{c}
\phi^{e}_{+} \\ 
\phi^{h}_{-} \\ 
\phi^{e}_{-} \\ 
\phi^{h}_{+}
\end{array}
\right) ,
\label{eq:BdGD_arm}
\end{eqnarray}
with $E$ positive unless otherwise specified. Each valley Hamiltonian is
written as $\hat{H}_{\pm }=-i\hbar v\left[ \pm \partial _{x}\hat{\sigma}_{x}+\partial_{y}\hat{\sigma}_{y}\right]$. Two-dimensional spinors $\phi^{e,h}_{\pm}$ represent electron or hole like excitations for each valley in lattice space. Whenever the pairing potential is assumed constant, the low energy spectrum is given by $E=\sqrt{\Delta ^{2}+\left(E_{F}^{S}-\hbar v\sqrt{k^{2}+q^{2}}\right) ^{2}}$. We define the transversal momentum as $\hbar vk_{e,h}^S = \sqrt{\left( E_{F}^{S}\pm \Omega \right)
^{2}-\left( \hbar vq\right)^{2}}$, with $\Omega =\sqrt{E^{2}-\Delta ^{2}}$. The solutions of the BdGD equation are a direct product of graphene's bulk
solutions and the usual BCS solutions. The pair potential couples electron and
hole-like excitations with opposite momentum, which means different
valley index, thus $\phi^{e}$ and $\phi^{h}$ correspond to different valleys. This allows us to reduce the degrees of freedom of the problem to Nambu and pseudospin spaces. 

In order to study a semi-infinite superconducting graphene region with an armchair edge extending from $x_L=0$ to $x_R \rightarrow \infty $, we define a finite and constant pair potential $\Delta(x)=\Theta (x)\Delta$. Reflection amplitudes are then fixed to $r^{+}_L=r^{-}_L=\mp 1 \equiv r$ and $r^{+}_R=r^{-}_R=0$. The asymptotic solutions are
\begin{eqnarray}
\psi_{<}^{e(h)+}\!\left( x\right)\! \!&=&\!\! \left\{
e^{\mp ik_{e(h)}x}\!\!\left( 
\begin{array}{c}
\!\!\varphi_{2e(1h)}\!\! \\ 
0
\end{array}
\!\!\right)\! +\! r e^{\pm ik_{e(h)}x}\!\!\left( 
\begin{array}{c}
0 \\ 
\!\!\varphi_{2e(1h)}\!\!
\end{array}
\!\!\right)  \right\} \nonumber \\
&&\!\otimes \left( \begin{array}{c}
u (v) \\ 
v (u)
\end{array}\right) \\
\psi_{>}^{e(h)+}\!\left( x\right)\! \!&=&\!\! e^{\pm ik_{e(h)}x}\!\!\left( 
\begin{array}{c}
\varphi_{1e(2h)} \\ 
0
\end{array}
\right) \otimes \left( \begin{array}{c}
u (v) \\ 
v (u)
\end{array}\right) \\
\psi_{<}^{e(h)-}\!\left( x\right)\! \!&=&\!\! \left\{ e^{\mp ik_{e(h)}x}\!\!\left( 
\begin{array}{c}
0 \\ 
\!\!\varphi_{1e(2h)}\!\!
\end{array}
\right)\! +\! r e^{\pm ik_{e(h)}x}\!\left( 
\begin{array}{c}
\!\!\varphi_{1e(2h)}\!\! \\ 
0
\end{array}
\!\right) \right\} \nonumber \\
&&\! \otimes \left( \begin{array}{c}
u (v) \\ 
v (u)
\end{array}\right) \\
\psi_{>}^{e (h)-}\!\left( x\right)\! \!&=&\! e^{\pm ik_{e(h)}x}\!\left( 
\begin{array}{c}
0 \\ 
\!\!\varphi_{2e(1h)}\!\!
\end{array}
\right) \otimes \left( \begin{array}{c}
u (v) \\ 
v (u)
\end{array}\right) ,
\end{eqnarray}
where $\varphi_{ie(h)}$, with $i=1,2$, is obtained from Eq. (\ref{eq:basis})  making $\alpha \rightarrow \alpha^S_{e\left( h\right) }$ and defining $e^{i\alpha^S_{e\left( h\right)}}=\hbar v\left( k^S_{e(h)}+iq\right) /|E_{F}^{S}\pm \Omega |$ and $u^{2}\left( v^{2}\right) =\left( 1\pm \Omega /E\right) /2$. The function $\psi_{<}^{e(h)±}$ describes an incoming electron-like (hole-like) quasiparticle with energy $E>E_F^S$ ($E<E_{F}^S$) and valley index $\pm$ that is reflected at $x_L$ into an electron-like (hole-like) quasiparticle with valley index $\mp$.

The $8 \times 8$ matrix Green's function can be written as the superposition of  all possible quasi-electron and quasi-hole injection processes depicted in Figs. \ref{fig:solutions_ribbon}(b) and \ref{fig:solutions_ribbon}(c). The resulting Green function, written in Nambu, valley and pseudospin spaces, has the form
\begin{equation}
\begin{array}{c}
\check{\hat{G}}_{\psi}^{S}\left( x<x^{\prime }\right) = \\
\left\{ 
\begin{array}{cr}
\displaystyle\sum_{\substack{{j = e,h}\\{  \mu ,\nu = +,-}}} A_{\mu \nu }^{j} \psi^{j \mu}_{<}\left( x\right) \cdot \psi^{j \nu T}_{>} \left( x^{\prime }\right) \left( \check{\gamma} \otimes \hat{\tau}_0 \right)& ;x<x^{\prime } \\ 
\displaystyle\sum_{\substack{{j=e,h}\\{\mu ,\nu =+,- }}} A_{\mu \nu}^{\prime j} \psi^{j \mu}_{>}\left( x\right) \cdot \psi^{j \nu T}_{<} \left( x^{\prime }\right) \left( \check{\gamma} \otimes \hat{\tau}_0 \right) & ;x>x^{\prime }
\end{array} \right. ,
\end{array}
\end{equation}
where $\tau_{x,y,z}$ and $\tau_0$ are Pauli matrices in Nambu space. The parity matrix $\check{\gamma}$ is diagonal in this space and only acts on graphene's valley and pseudospin degrees of freedom.

As in the normal case, the coefficients $A^{(\prime)j}_{\mu,\nu}$ with $\mu=\nu$ vanish as they correspond to processes which change the valley polarization without reflection at the edges. Additionally, the processes that couple quasi-electron with quasi-hole and vice versa have zero probability. This can be explicitly found from the condition given by Eq. (\ref{eq:BC_AB_Dirac}) where $A^{(\prime)j}_{\mu = \nu} =0$, $A^{e}_{+-(-+)}=A^{\prime e}_{-+(+-)}=iEe^{+(-) i\alpha_e}/\left( 2\hbar v\Omega \cos{\alpha_e}\right) $ and $A^{h}_{-+(+-)}=A^{\prime h}_{+-(-+)}=iEe^{+(-) i\alpha_h}/\left( 2\hbar v\Omega \cos{\alpha_h}\right) $. Thus, after applying boundary conditions and solving Eq. (\ref{eq:BC_AB_Dirac}), the valley superposed Green function in Nambu space for a superconducting region with an armchair edge is 
\begin{equation}
\begin{array}{ccl}
\check{G}_{\phi}^{S \text{,arm}}\left( x,x^{\prime }\right) &=& \frac{-i}{2\hbar v} \left\{ \left[ \hat{A}^S_e + \hat{A}^S_h \right] \otimes \left( 
\frac{E}{\Omega } \hat{\tau}_0 + \frac{\Delta}{\Omega } \hat{\tau}_x
\right) \right. \\
&& \left. + \left[ \hat{A}^S_e - \hat{A}^S_h \right] \otimes \hat{\tau}_z \right\}
\end{array}
\label{eq:FdG_arm_super} ,
\end{equation}
with the following definitions
\begin{eqnarray}
\hat{A}^S_{e(h)} &=& f_{e(h)} \left( \cos^{-1} {\alpha^S_{e(h)}} \hat{\sigma}_0 + \tan{ \alpha^S_{e(h)}} \hat{\sigma}_y \right) \pm ig_{e(h)} \hat{\sigma}_x \nonumber \\
f_{e(h)} &=&\cos {\left[ K\left( x'-x\right) \right] }e^{\pm
ik^S_{e(h)}\left\vert x'-x\right\vert } \nonumber \\
&& + r \cos {\left[ K\left(
x'+x\right) \right] }e^{\pm ik^S_{e(h)}\left( x'+x\right) }
\nonumber \\
g_{e(h)} &=&\mbox{sgn}\left(x'-x\right)\sin {\left[ K\left( x'-x\right) \right] }e^{\pm
ik^S_{e(h)}\left\vert x'-x\right\vert } \nonumber \\
&& + r \sin {\left[ K\left(
x'+x\right) \right] }e^{\pm ik^S_{e(h)}\left( x'+x\right) } . \nonumber
\end{eqnarray}

It is interesting to take this expression to the heavily doped limit in which $E_F^S \gg E, \Delta, \hbar vq$ and thus $\cos{\alpha^S_{e,h}} \rightarrow 1$ and $\sin{\alpha^S_{e,h}} \rightarrow 0$. Within this approximation, for a microscopic distance $x_0 \sim a$ from the interface we have
\begin{equation}
\begin{array}{c}
\check{G}_{\phi}^{S \text{,arm}}\left( x_0,x_0\right) \approx  \frac{-i}{\hbar v} \left\{ \left[ 1 + r \cos{2Kx_0} \right] \hat{\sigma}_0 \right. \\   
\left. \otimes \left( 
\frac{E}{\Omega } \hat{\tau}_0 + \frac{\Delta}{\Omega } \hat{\tau}_x
\right) + i r \sin{2Kx_0} \hat{\sigma}_x \otimes \hat{\tau}_z \right\} .
\end{array}
\label{eq:FdG_semi_HDSC}
\end{equation}

For the boundary condition of vanishing the derivative of the wave function ($r=1$), we can evaluate the Green function at the edge ($x_0=0$). Thus,
\begin{equation}
\check{G}_{\phi}^{S \text{,arm}}\left( 0,0 \right) \approx - \frac{2i}{\hbar v \Omega} \left(E \hat{\sigma}_0 \otimes \hat{\tau}_0 + \Delta \hat{\sigma}_0 \otimes \hat{\tau}_x \right) ,
\label{eq:FdG_semi_BCS}
\end{equation}
which corresponds to the momentum independent BCS Green function without sublattice structure \cite{burset08}. Then, $2/\hbar v$ corresponds to an averaged Fermi energy DOS per unit length for the superconductor.

On the other hand, for the boundary condition over the wave function ($r=-1$) the Green function vanishes at $x_0=0$. For finite $x_0$, the Green function can exhibit a structure in pseudospin space given by the last term in Eq. (\ref{eq:FdG_semi_HDSC}). This expression is used in the following sections when we microscopically couple normal and superconducting regions.

\subsection{Zig-zag edges}

The geometry of the graphene sheet is now set to have a zigzag edge along
the $y$-axis. With this new orientation of the layer the Brillouin zone rotates as illustrated in the upper part of the right panel of Fig. \ref{fig:bulk_edges}, which allows to select the Dirac points at $\mathbf{K}_{\pm} = (0,\pm K)$. We still call $\hbar q$ the conserved momentum along the $y$ direction. We consider either a finite layer with edges at positions $x_{L}$ and $x_{R}$ or a semi-infinite one in which $x_L$ or $x_R$ goes to infinity. Zigzag edges are formed by a line of atoms all pertaining to one sublattice. These edges do not mix valleys so we can treat them separately and use a Dirac Hamiltonian of just one valley, $\hat{H}_{+} = \hbar v\left( k\hat{\sigma}_{x} + q\hat{\sigma}_{y}\right) - E_{F}\hat{\sigma}_{0}$. Thus, the asymptotic wave functions including a reflection at the edge are 
\begin{eqnarray}
\phi_{<}(x) &=& e^{-ikx}\varphi _{2}+r_{L}e^{ikx}\varphi_{1}
\\
\phi_{>}(x) &=& e^{ikx}\varphi _{1}+r_{R}e^{-ikx}\varphi_{2}.
\end{eqnarray}

The reflection amplitudes depend on the atoms chosen to form the zigzag
edge. Thus, if we choose the border at $x_{L}$ to be formed by atoms of $A$ ($B$) lattice, the one at $x_{R}$ is formed by atoms pertaining to $B$ ($A$) lattice (see right panel of Fig. \ref{fig:bulk_edges}). Imposing the boundary conditions $\left. \phi _{<}(x_{L}) \right\vert_{A(B)} = \left. \phi _{>}(x_{R})\right\vert_{B(A)}=0$ we obtain the reflection amplitudes: 
\begin{equation}
r_{L(R)}^{A} = -e^{\pm i\alpha }e^{\mp 2ikx_{L(R)}} \quad
r_{L(R)}^{B} = e^{\mp i\alpha }e^{\mp 2ikx_{L(R)}}.
\label{eq:zz_ref_coef}
\end{equation}

When combining the asymptotic solutions to build the Green functions, as it was done in the previous section, the elements of the full Green function that mix valleys ($+-,-+$) are zero. In what follows the full Green function is understood to represent only a projection over one valley (i.e. $\check{G}_{\psi}^{\text{zz,++}}$). Following the steps given in the previous sections we can write
\begin{equation}
\hat{G}_{\psi}^{\text{zz}}(x,x^{\prime })=\left\{ 
\begin{array}{cr}
A\phi _{<}(x)\cdot \phi_{>}^{T}(x^{\prime })\cdot \hat{\gamma} & 
;x<x^{\prime } \\ 
A^{\prime }\phi _{>}(x)\cdot \phi_{<}^{T}(x^{\prime })\cdot 
\hat{\gamma} & ;x>x^{\prime }
\end{array}
\right. .
\end{equation}
When restricted to a single valley, the parity matrix $\hat{\gamma}$ is equivalent to $\hat{\sigma}_{z}$ (see Appendix \ref{appendix_A} for a detailed discussion). For a ribbon of graphene with the left zigzag edge of type $B$ (and thus the right one of type $A$) we have
\begin{equation}
\begin{aligned}
&\hat{G}_{\psi}^{\text{zz}}(x,x^{\prime}) = \frac{-i}{2\hbar v\cos{\alpha} \left(
1-r_{L}^{B}r_{R}^{A}\right) } \\
&\times \left\{ 
\begin{array}{cr}
\begin{array}{c}
e^{ik (x'-x) } \varphi _{2}\varphi _{2}^{\dagger
}+r_{L}^{B}r_{R}^{A} e^{-ik (x'-x) } \varphi_{1}\varphi_{1}^{\dagger } + \\ +r_{L}^{B}e^{ik(x'+x)}\varphi _{1}\varphi
_{2}^{\dagger }+r_{R}^{A}e^{-ik(x'+x)}\varphi _{2}\varphi _{1}^{\dagger }
\end{array} & 
;x<x^{\prime } \\
& \\ 
\begin{array}{c}
e^{ik(x-x') } \varphi_{1} \varphi_{1}^{\dagger} + r_{L}^{B} r_{R}^{A} e^{-ik(x-x') } \varphi_{2} \varphi_{2}^{\dagger} + \\
 + r_{L}^{B}e^{ik(x+x')}\varphi _{1}\varphi_{2}^{\dagger} + r_{R}^{A} e^{-ik(x+x')} \varphi_{2} \varphi_{1}^{\dagger }
\end{array} & 
;x>x^{\prime }
\end{array} \right.
\end{aligned}
\label{eq:FdG_zz}
\end{equation}
For the other valley we obtain the same result with the change $\varphi_{1,2} \leftrightarrow \varphi_{2,1}$. As it was explained in the previous section, $\cos {\alpha }=0$ gives the bulk dispersion relation. Furthermore, if the layer has a finite width which is set to $W=x_{R}-x_{L}$, the condition $1 - r_{L}^{B}r_{R}^{A} = 0$ transforms into 
\begin{equation}
e^{2ikW}=\frac{q+ik}{q-ik} .
\label{eq:disp_rel_zz}
\end{equation}
For $k$ real, this expression leads to the quantization of transverse
momentum in the ribbon (i.e. $q=k_{n}/\tan {k_{n}W}$). On the other hand, if
the transverse momentum is a pure imaginary number ($k=-iz$) Eq. (\ref{eq:disp_rel_zz}) transforms into 
\begin{equation}
e^{-2zW}=\frac{q-z}{q+z},
\end{equation}
whose solutions correspond to surface states, at $E=E_F$, localized along the edges of the ribbon \cite{brey06}.

Analogously to the armchair case, we now consider a superconducting region spread over the $x>0$ infinite half-plane with a zigzag edge at $x=0$. If we couple electronic excitations with $\mathbf{K}_+$ index to hole-like quasiparticles with the other valley index, Eq. (\ref{eq:BdGD_arm}) is reduced to a $4 \times 4$ matrix equation in Nambu and pseudospin spaces
\begin{equation}
\left( 
\begin{array}{cc}
\hat{H}_{+}-E_{F}^{S} & \Delta \\ 
\Delta^{*} & E_{F}^{S}-\hat{H}_{+}
\end{array}
\right) \left( 
\begin{array}{c}
\phi^{e}_+ \\ 
\phi^{h}_-
\end{array}
\right) =E\left( 
\begin{array}{c}
\phi^{e}_+ \\ 
\phi^{h}_-
\end{array}
\right) .
\label{eq:BdGD_zz}
\end{equation}

The asymptotic wave functions for quasi-electron and quasi-hole injection are written in Nambu and pseudospin space as
\begin{eqnarray}
\phi_{<}^{e(h)}\left( x\right) &=& \left\{ e^{\mp ikx}\varphi _{2e(1h)}  \right. \nonumber \\ 
 &+& \left. r_{L}^{B,e(h)} e^{\pm ikx}\varphi _{1e(2h)} \right\} \otimes \left( 
\begin{array}{c}
u\left( v\right) \\ 
v\left( u\right)
\end{array}
\right)  \\
\phi_{>}^{e(h)}\left( x\right) &=& e^{\pm ikx}\varphi _{1e(2h)} \otimes \left( 
\begin{array}{c}
u\left( v\right) \\ 
v\left( u\right)
\end{array}
\right) ,
\end{eqnarray}
where the edge has been chosen to be formed by atoms from sublattice $B$. Boundary conditions for the wave function at the zigzag edge determine the
reflection amplitudes. Thus, for electronic excitations we substitute $\alpha \rightarrow \alpha_e^S$ in Eq. 
\eqref{eq:zz_ref_coef}, while we change $\alpha \rightarrow -\alpha^S_{h}$ for hole excitations. The resulting Green's function for a semi-infinite superconducting region reads
\begin{equation}
\begin{array}{ccl}
\check{G}^{S,\text{zz}}_{\psi}(x,x') &=& \frac{-i}{2\hbar v} \left\{ \left[ \hat{Z}^S_e + \hat{Z}^S_h \right] \otimes \frac{1}{\Omega} \left( E \hat{\tau}_0 + \Delta \hat{\tau}_x \right)  \right. \\
&& \left. + \left[ \hat{Z}^S_e - \hat{Z}^S_h \right] \otimes \hat{\tau}_z \right\} ,
\label{eq:FdG_zz_super}
\end{array}
\end{equation}
where
\begin{equation}
\begin{array}{ccl}
\hat{Z}^S_{e(h)} &=& \frac{e^{\pm ik^S_{e(h)} (x+x')}}{2\cos {\alpha^S_{e(h)}}} e^{\mp i \alpha^S_{e(h)}} \varphi_{1e(2h)} \varphi_{2e(1h)}^{\dagger} \\
&&+\frac{e^{\pm i k^S_{e(h)} \left\vert x'-x \right\vert }}{2\cos{\alpha^S_{e(h)}}} \left\{ \begin{array}{cr}
\varphi_{2e(1h)} \varphi_{2e(1h)}^{\dagger} & ;x<x' \\
\varphi_{1e(2h)} \varphi_{1e(2h)}^{\dagger} & ;x>x'
\end{array}\right. . \nonumber
\end{array}
\end{equation}
When Eq. (\ref{eq:FdG_zz_super}) is evaluated at the edge of the graphene layer it reduces to
\begin{equation}
\begin{array}{c}
\check{G}^{S,\text{zz}}_{\psi}(0,0) = 
\frac{-i}{2\hbar v} 
\left\{ 
\left( \begin{array}{cc}
 e^{-i\alpha^S_{e}} + e^{i\alpha^S_{h}} & 0 \\ 0 & 0
\end{array} \right) \right. \\
\left. \otimes  \frac{1}{\Omega} \left( E \hat{\tau}_0 + \Delta \hat{\tau}_x \right) + \left( \begin{array}{cc}
 e^{-i\alpha^S_{e}} - e^{i\alpha^S_{h}} & 0 \\ 0 & 0
\end{array} \right) \otimes \hat{\tau}_z 
\right\}  
\\
+ \frac{-i}{2\hbar v} \left\{ \begin{array}{cr}
\left( \begin{array}{cc}
0 & -1 \\ 0 & 0
\end{array} \right) \otimes \hat{\tau}_z & ; 0<x<x' \\
\left( \begin{array}{cc}
0 & 0 \\ 1 & 0
\end{array} \right) \otimes \hat{\tau}_z & ; 0<x<x'
\end{array} \right.
\end{array}
\end{equation}

Furthermore, in the heavily doped limit ($E_F^S \gg E, \Delta, \hbar v q$), the previous expression reduces to
\begin{equation}
\begin{array}{ccl}
\check{G}^{S,\text{zz}}_{\psi}(0,0) &\approx & \frac{-i}{2\hbar v} \left\{ \left( \hat{\sigma}_{0} + \hat{\sigma}_z \right) \otimes  \frac{1}{\Omega} \left( E \hat{\tau}_0 + \Delta \hat{\tau}_x \right) \right. \\
&& \left. \mp \frac{1}{2} \left( \hat{\sigma}_{x} \pm i \hat{\sigma}_y \right) \otimes \hat{\tau}_z \right\} ,
\end{array}
\label{eq:FdG_semi_zz}
\end{equation}
where the sign identifies the cases with $x<x'$ or $x>x'$. Thus, in this limit we roughly have a BCS Green function projected to the site $A$.

\section{Dyson equation for coupling separate regions}

Once the expressions for bulk, armchair and zigzag graphene sheets,
as well as for the semi-infinite superconducting regions have been
obtained, we concentrate in the determination of the Green functions for the
coupled system. 
This amounts in principle to solve the integral Dyson equation

\begin{equation}
\begin{aligned}
\hat{G}_{\phi}&(x,x^{\prime}) = \hat{g}_{\phi}(x,x^{\prime}) \\
&+\int dx_1 dx_2 \hat{g}_{\phi}(x,x_1)\hat{V}(x_1,x_2) \hat{G}_{\phi}(x_2,x^{\prime}),
\end{aligned}
\label{eq:int_dyson}
\end{equation} 
where now $\hat{G}_{\phi}$ denotes the valley superposed Green function of the coupled system, $\hat{g}_{\phi}$ corresponds to the uncoupled ones and $\hat{V}$
is an appropriate perturbation describing the coupling between
the two regions. In general, we must use valley superposed Green functions in Dyson's equation to have a microscopic description of the coupling between graphene layers. In order to simplify the model while still keeping the
possibility to describe interfaces of arbitrary transparency we shall 
assume that the perturbation only connect points on an atomic scale
close to the interface. We thus define a general perturbation
\begin{equation}
\hat{V} \left( x_1 , x_2 \right) = \beta t_g a_{eff} \delta \left( x_1 - x_0 \right)
 \delta \left( x_2 + x_0 \right) \hat{\sigma}_x ,
\label{eq:perturbation}
\end{equation}
where $\beta \in \left[0,1 \right]$ is a dimensionless parameter that controls the transparency of the interface and $x_0$ is a microscopic distance from the interface. The effective hopping matrix $t_g a_{eff} \hat{\sigma}_x$ has a structure dictated by the change of sublattice associated with the hopping between the two graphene sheets at the interface. In this expression $a_{eff}$ represents an ``effective distance" which we determine from the condition that the bulk graphene result is recovered when coupling two semi-infinite regions with $\beta=1$. 

We first discuss the case of two semi-infinite graphene sheets with
armchair edges. The simplest possible choice would be $x_0 \rightarrow 0^+$, which can be applied when the condition of vanishing the derivative of the wave function at the edge has been taken. 
In this way one may convert Eq. (\ref{eq:int_dyson}) into an algebraic equation, which for the local Green functions at the interface can be written as
\begin{equation}
\hat{G}_{R(L)}=
\left[ \hat{g}_{R(L)}^{-1}- \left( \beta t_g a_{eff} \right)^2 \hat{\sigma}_x 
\hat{g}_{L(R)} \hat{\sigma}_x \right] ^{-1},  
\label{eq:dyson-algebraic}
\end{equation}
where $R,L$ denotes  $(x,x^{\prime}) \rightarrow (0^{\pm},0^{\pm})$ 
and $\hat{g}_{L(R)}$ are given by
\begin{equation}
\hat{g}_{L(R)}=\frac{-2i}{\hbar v}\left( \cos
^{-1}{\alpha }\hat{\sigma}_{0}+\tan {\alpha }\hat{\sigma}_{y}\right).
\end{equation}

One can check that the bulk Green function is recovered for $a_{eff}= \sqrt{3}a/4$, which therefore sets the condition of perfect transparency for this case. The reader is addressed to Appendix \ref{appendix_B} for more details about the bulk results for a normal or superconducting graphene region.

A slightly more cumbersome matching condition has to be introduced if
one would like to reproduce the limiting results of the TB model for such an interface \cite{burset08}.
This requires using the Green functions obtained by vanishing the 
wave function at the edges of the graphene sheet and take
$x_0 = a/4$. 
One obtains then a Dyson equation like (\ref{eq:dyson-algebraic})
with $R,L$ denoting  $(x,x^{\prime})=(\pm\frac{a}{4},\pm\frac{a}{4})$ 
and $\hat{g}_{L,R}$ given by
\begin{equation}
\hat{g}_{L(R)}=\frac{-\sqrt{3}i}{2\hbar v}\left[ \sqrt{3
}\left( \cos ^{-1}{\alpha }\hat{\sigma}_{0}+\tan {\alpha }\hat{\sigma}
_{y}\right) - i\hat{\sigma}_{x}\right] .
\label{eq:TB_arm_semi}
\end{equation}

These expressions exactly reproduce the TB results in the continuum
limit, this is $a\rightarrow 0$ but with $Ka \rightarrow 4\pi/3$ \cite{burset09}.
One can further check that for this case the bulk graphene result is recovered
for $a_{eff}= a/2$ which is the distance between lines of atoms along the $x$ direction in an armchair edge.

We now focus on zigzag edges. In this case one can disregard the valley
mixing but care should be taken due to the discontinuity of the Green
functions at $x=x^{\prime}$. As discussed in the previous section we select the condition of vanishing one of the wave function components at the interface and take $x_0 \rightarrow 0^+$. The precise value of $x_0 \sim a$ is here irrelevant because the Dirac points lie along the $y$ direction.

When we evaluate Eq. (\ref{eq:FdG_zz}) at the edge of each region we get
\begin{eqnarray}
\hat{g}_{L} &=&\frac{-i}{\hbar v}\left\{ 
\begin{array}{ccr}
\left( 
\begin{array}{cc}
0 & -1 \\ 
0 & e^{-i\alpha }
\end{array}
\right) & \equiv \hat{g}^+_{L} & ;x<x^{\prime } \rightarrow 0^-\\ 
\left( 
\begin{array}{cc}
0 & 0 \\ 
1 & e^{-i\alpha }
\end{array}
\right) & \equiv \hat{g}^-_{L} & ;x^{\prime }<x \rightarrow 0^-
\end{array}
\right. \\
\hat{g}_{R} &=&\frac{-i}{\hbar v}\left\{ 
\begin{array}{ccr}
\left( 
\begin{array}{cc}
e^{-i\alpha } & -1 \\ 
0 & 0
\end{array}
\right) & \equiv \hat{g}^+_{R} & ;x<x^{\prime }\rightarrow 0^+ \\ 
\left( 
\begin{array}{cc}
e^{-i\alpha } & 0 \\ 
1 & 0
\end{array}
\right) & \equiv \hat{g}^-_{R} & ;x^{\prime }<x\rightarrow 0^+
\end{array}
\right.
\end{eqnarray}
where we have implicitly assumed a $A(B)$ termination for the $L(R)$ 
graphene sheet. The Green function of the coupled system 
must be constructed including the discontinuity of the uncoupled
Green's functions at the edges, i.e for $x,x^{\prime} < 0$
\begin{equation}
\begin{array}{c}
\hat{G}_L \left( x,x^{\prime }\right) =\hat{g}\left(
x,x^{\prime }\right) +\left(t_g a_{eff}\right)^2 \hat{g}\left( x,0^-\right) \hat{\sigma}_x \\ 
\times \left[ 1-\left(t_g a_{eff}\right)^2 \hat{g
}_{R}^- \hat{\sigma}_x \hat{g}_{L}^+
\hat{\sigma}_x \right] ^{-1}\hat{g}_{R}^- \hat{\sigma}_x \hat{g}\left( 0^{-},x^{\prime }\right) ,
\end{array}
\end{equation}
where $\hat{g}\left( x,0^-\right) =\exp \left( -ikx\right) \hat{g
}_{L}^+ $, $\hat{g}_{L}\left( 0^{-},x^{\prime
}\right) =\exp \left( -ikx^{\prime }\right) \hat{g}_{L}^-$ and $\hat{g}\left(
x,x^{\prime }\right)$ corresponds to Eq. (\ref{eq:FdG_zz}) with $r_L^B=0$. When $x<x^{\prime }$, and using that for this case $r_{L}^{B}=-r_{R}^{A}$, this results in
\begin{equation}
\begin{array}{c}
\hat{G}_L \left( x,x^{\prime }\right) =\frac{-i}{2\hbar v\cos {
\alpha }}\left[ e^{ik\left( x-x^{\prime }\right) }\varphi _{2}\varphi
_{2}^{\dagger } \right. \\
\left. +e^{-ik\left( x+x^{\prime }\right) }r_{R}^{A}\varphi
_{2}\varphi _{1}^{\dagger }\left( 1-\frac{\beta^{2}\left( 1+e^{-2i\alpha
}\right) }{\frac{3a^2}{4a_{eff}^2}+\beta^{2}e^{-2i\alpha }}\right) \right] ,
\end{array}
\end{equation}
where we have replaced $t_g = 2 \hbar v /(\sqrt{3} a)$. The bulk result is recovered at perfect transparency ($\beta=1$) for $a_{eff}=\sqrt{3}a/2$, which corresponds to the horizontal separation between atoms of the same sublattice. This result corresponds to the projection over the point $\mathbf{K}_+$. For the other valley, the same result with $\varphi_1 \leftrightarrow \varphi_2$ is reached and the bulk result is recovered for perfect transparency superposing both valleys.

\section{Graphene-superconductor junction}

All the ingredients necessary to analyze the graphene-superconductor coupling have been already introduced. In the present section, we analytically derive the Green function of the coupled system both for armchair and zigzag edges. We locate the interface at $x=0$, parallel to the $y$ axis, with the normal region covering the $x<0$ infinite half-plane. Since the superconducting region is represented in Nambu space, in order to use Dyson's equation as has been explained in section III, the Green's functions of the normal region are written as 
\begin{equation}
\check{g}_L=\left( 
\begin{array}{cc}
\hat{g}_e & 0 \\ 
0 & \hat{g}_h
\end{array}
\right).
\label{eq:FdG_nambu}
\end{equation}

Electron excitations are states above the Fermi energy. Then, $\hat{g}_e$ corresponds to Eq. (\ref{eq:FdG_arm_f}) for armchair edges or Eq. (\ref{eq:FdG_zz}) for zigzag edges if we set $\alpha_e = \alpha$, with $\hbar v k_e = \sqrt{\left( E_F+E \right)^2 - \left( \hbar v q \right)^2}$. On the other hand, hole excitations have energies below $E_F$. Thus, $\hat{g}_h$ corresponds to the same equations but with the change $\alpha_h = \arcsin{\hbar v q / \left(E_F-E \right)}$, with $\hbar v k_h = \sqrt{\left( E_F-E \right)^2 - \left( \hbar v q \right)^2}$. For the superconducting region ($R$), Eq. (\ref{eq:FdG_arm_super}) for armchair edges or Eq. (\ref{eq:FdG_zz_super}) for zigzag edges, evaluated at the interface, stand for $\check{g}_R$ in Dyson's equation. 

The coupling with the superconductor appears in the Green function as a perturbation of the semi-infinite result $\check{g}_L$ given by Eqs. (\ref{eq:FdG_arm_f}) and (\ref{eq:FdG_zz}). Thus we can write
\begin{equation}
\check{G}_{L}(x,x^{\prime}) = \check{g}_L (x,x^{\prime}) + \delta\check{G}(x,x^{\prime}),
\end{equation}
where $\delta\check{G}$ is the correction with respect to the uncoupled case. Our formalism allows in principle to obtain analytical results for $\delta \check{G}$ in the case of arbitrary effective hopping and coordinates $x$ and $x'$. 

For the case of an armchair interface, using the boundary conditions of vanishing the derivative of the wave function and setting $\beta = 1$, for $x=x'$ we obtain
\begin{equation}
\begin{array}{c}
\delta \hat{G}_{ee}^{\text{arm}}(x=x') = \frac{i}{\hbar v}\frac{e^{-2ik_{e}x}}{\cos{\alpha_e}} \\
\times \left\{ \left( 
\begin{array}{cc}
\cos 2Kx & -i\sin \left( 2Kx+\alpha_e \right) \\ 
-i\sin \left( 2Kx-\alpha_e \right) & \cos 2Kx
\end{array}
\right) \right. \\
\left. + \frac{N_{\text{arm}}}{D_{\text{arm}}} \left( 
\begin{array}{cc}
\sin \alpha_e & -i \\ 
i & \sin \alpha_e
\end{array}
\right) \right\},
\end{array}
\end{equation}
where we have defined the auxiliary quantities 
\begin{eqnarray}
N_{\text{arm}} &=& 2\sin{\alpha_e} \left( 1 + \frac{E}{\Omega} \cos{\alpha_h} \right) + \frac{\Delta^2}{\Omega^2} \left( \sin{\alpha_e} - \sin{\alpha_h} \right) \nonumber \\
D_{\text{arm}} &=& 2 + 2\cos \alpha_e \cos \alpha_h + 2\frac{E}{\Omega }\left( \cos \alpha_e + \cos \alpha_h \right) \nonumber \\
&& + \frac{\Delta ^{2}}{\Omega ^{2}}\left( 1+ \cos{\left[ \alpha_e +
\alpha_h \right]} \right)  \nonumber .
\end{eqnarray}

The denominator $D_{\text{arm}}$ contains information on the spectral properties of the coupled system. In particular, the condition $D_{\text{arm}}=0$ is satisfied by states with energy given by
\begin{equation}
\frac{E}{\Delta} = \pm 
\frac{\left(e^{(\lambda_e+\lambda_h)/2}-\mbox{sign}(E^2-E_F^2)
e^{-(\lambda_e+\lambda_h)/2}\right)}{2\sqrt{\cosh{\lambda_e}\cosh{\lambda_h}}},
\label{eq:IBS_arm}
\end{equation}
where we define $\lambda_{e,h} = \mbox{sign}(q) \mbox{arcosh}(\hbar v q/|E\pm E_F|)$ for evanescent electron or hole states with $\left\vert \hbar v q \right\vert > \left\vert E_F \pm E \right\vert$. As demonstrated in Ref. \cite{burset09}, these states correspond to interface bound states (IBSs) arising from the interplay between normal and Andreev reflection at the junction region. The decay of these states into the normal region is given by $e^{x/\xi_{e,h}}$, with $\xi_{e,h}=\hbar v/\left( \left\vert E \pm E_F \right\vert \sinh \lambda_{e,h} \right)$, which can be clearly much larger than the BCS superconducting coherence length $\xi=\hbar v/\Delta$ when $E_F \ll \Delta$. In the opposite limit, $E_F \gg \Delta$, the states are confined to the interface.

On the other hand, for the calculation of the LDOS it is necessary to obtain the diagonal components of the local Green's function ($[ee,AA]$ and $[ee,BB]$). Whenever $Kx=n\pi$, they simplify to 
\begin{equation}
\begin{array}{c}
G_{L \vert ee,AA}^{\text{arm}}(x,x) = G_{L \vert ee,BB}^{\text{arm}}(x,x) = \\ \frac{-i}{\hbar v} \left\{ \frac{1}{\cos
\alpha_e } - e^{-2ik_{e}x}\frac{N_{\text{arm}}}{D_{\text{arm}}} \tan{\alpha_e}  \right\} .
\end{array}
\end{equation}

Furthermore, for analyzing the induced pairing correlations in the normal graphene layer, the electron-hole elements of $\delta \check{G}$ are needed. Within the same conditions of the previous result, but for any values of $x$ and $x'$, these elements reduce to
\begin{equation}
\begin{aligned}
\hat{G}&_{L \vert eh,AA(BB)}^{\text{arm}}(x,x') =
\frac{-i e^{-i k_{e} x} e^{i k_{h} x'}}{\hbar v D_{\text{arm}}}
\frac{ \Delta }{\Omega } \left\{ \cos \left[K\left(x-x' \right) \right] \right. \\
&\left. \times \left[1+\cos \left[ \alpha_e - \alpha_h\right] +\frac{E}{\Omega} \left( \cos \alpha_e + \cos \alpha_h \right) \right] \right.  \\ 
&\left. + \frac{1}{2} \sin \left[K\left(x-x' \right) \right] \frac{\cos \alpha_e -\cos \alpha_h}{\cos \alpha_e \cos \alpha_h} \right.\\ 
&\left. \times \left[ \sin \alpha_e + \sin \alpha_h + \frac{E}{\Omega} \sin \left[\alpha_e -\alpha_h \right]\right]
\right\} .
\end{aligned}
\end{equation}

We now analyze a normal-superconductor graphene junction with a zigzag edge at $x=0$. As it was explained at the beginning of this section, we have chosen the normal (superconducting) region to be at $x<0$ ($x>0$) and to have the edge formed by $A$ ($B$) atoms. For these conditions we obtain the following correction
\begin{equation}
\begin{array}{c}
\delta\check{G}_{ee}^{\text{zz}}(x,x') = \\ 
-\frac{i \beta^2 e^{-ik_e (x+x')}}{\hbar v D_{\text{zz}}}\left[ \frac{E}{\Omega} + \beta^2 \frac{e^{i\alpha_h}}{\cos{\alpha_h}} \right] \left( \begin{array}{cc}
1 & e^{-i\alpha_e} \\
-e^{-i\alpha_e} & -e^{-2i\alpha_e}
\end{array} \right) ,
\end{array}
\end{equation}
with $D_{\text{zz}}=1+\beta^2 \frac{E}{\Omega} \left( \frac{e^{-i\alpha_e}}{\cos{\alpha_e}} + \frac{e^{i\alpha_h}}{\cos{\alpha_h}} \right) + \beta^4 \frac{e^{-i\alpha_e} e^{i\alpha_h}}{\cos{\alpha_e}\cos{\alpha_h}}$. Again, when this denominator is set to zero one obtains the dispersion relation for the IBSs along this edge,
\begin{equation}
\frac{E}{\Delta} = \pm \frac{e^{(\lambda_e+\lambda_h)/2} - 
\beta^2 \mbox{sign}(E^2-E^2_F)
e^{-(\lambda_e+\lambda_h)/2}}
{\sqrt{(e^{\lambda_e} + \beta^2 e^{-\lambda_e}) 
(e^{\lambda_h} + \beta^2 e^{-\lambda_h})}} ,
\label{eq:IBS_zz}
\end{equation}
which is in agreement with the result of Ref. \cite{burset09}.

On the other hand, the electron-hole component of the Green function is given by
\begin{equation}
\begin{array}{c}
\check{G}_{L \vert eh}^{\text{zz}}(x,x') = \\ 
\frac{i \beta^2 e^{-ik_e x} e^{ik_h x'} }{\hbar v D_{\text{zz}}}\frac{\Delta}{\Omega} \left( \begin{array}{cc}
1 & e^{i\alpha_h} \\
-e^{-i\alpha_e} & -e^{-i\alpha_e} e^{i\alpha_h}
\end{array} \right) .
\end{array}
\end{equation}

\subsection{LDOS for armchair and zigzag edges}

\begin{figure*}
\includegraphics[width=7.5cm]{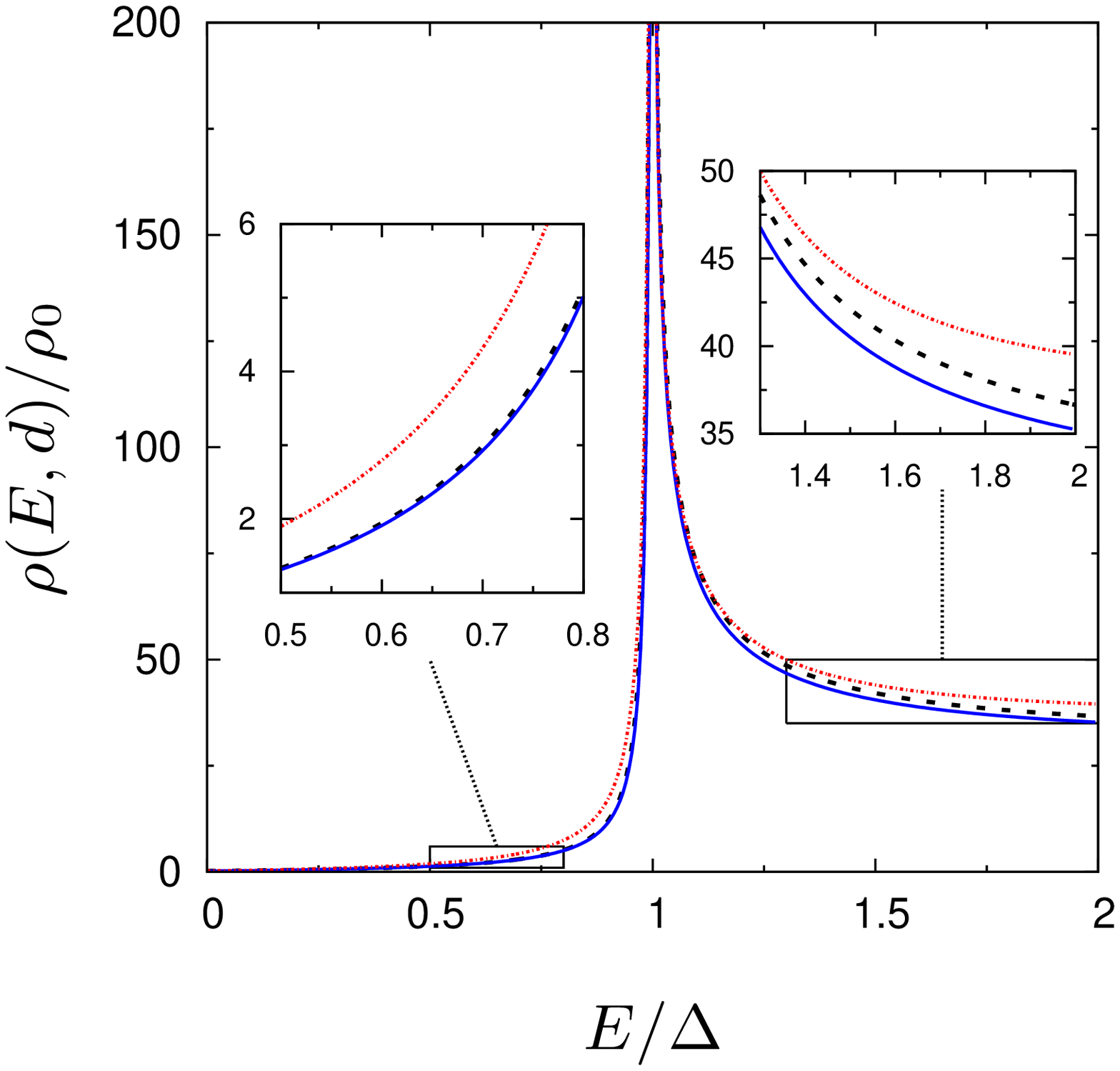}
\quad
\includegraphics[width=7.5cm]{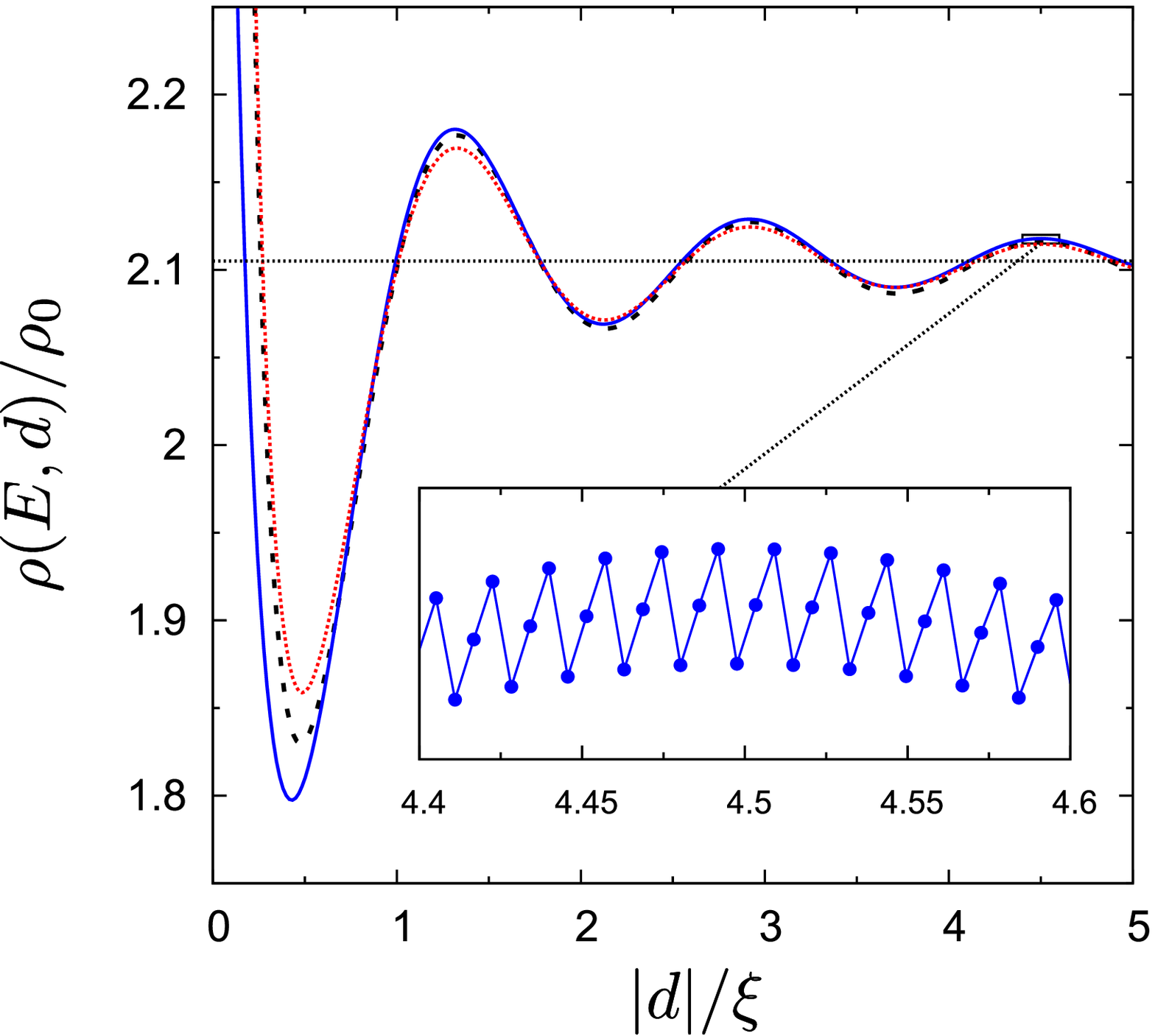}
\caption{(Color online) LDOS in the normal region with $E_F=0$ for a semi-infinite layer of graphene with armchair edges coupled to a superconducting region (left panel). Profiles for both boundary conditions (full blue line when vanishing the wave function and full red line for the derivative) are equivalent to the TB results (dashed line). The right panel shows the oscillating behavior of the LDOS for $E=2\Delta$ along the direction transversal to the interface. The transversal distance $d$ is normalized to the BCS superconducting coherence length $\xi$. All the results show the same oscillating behavior around the bulk value. The oscillatory behavior at the atomic scale is shown in the inset for the boundary condition of vanishing the wave function.}
\label{fig:LDOS}
\end{figure*}

\begin{figure*}
\centering
\includegraphics[width=8.cm]{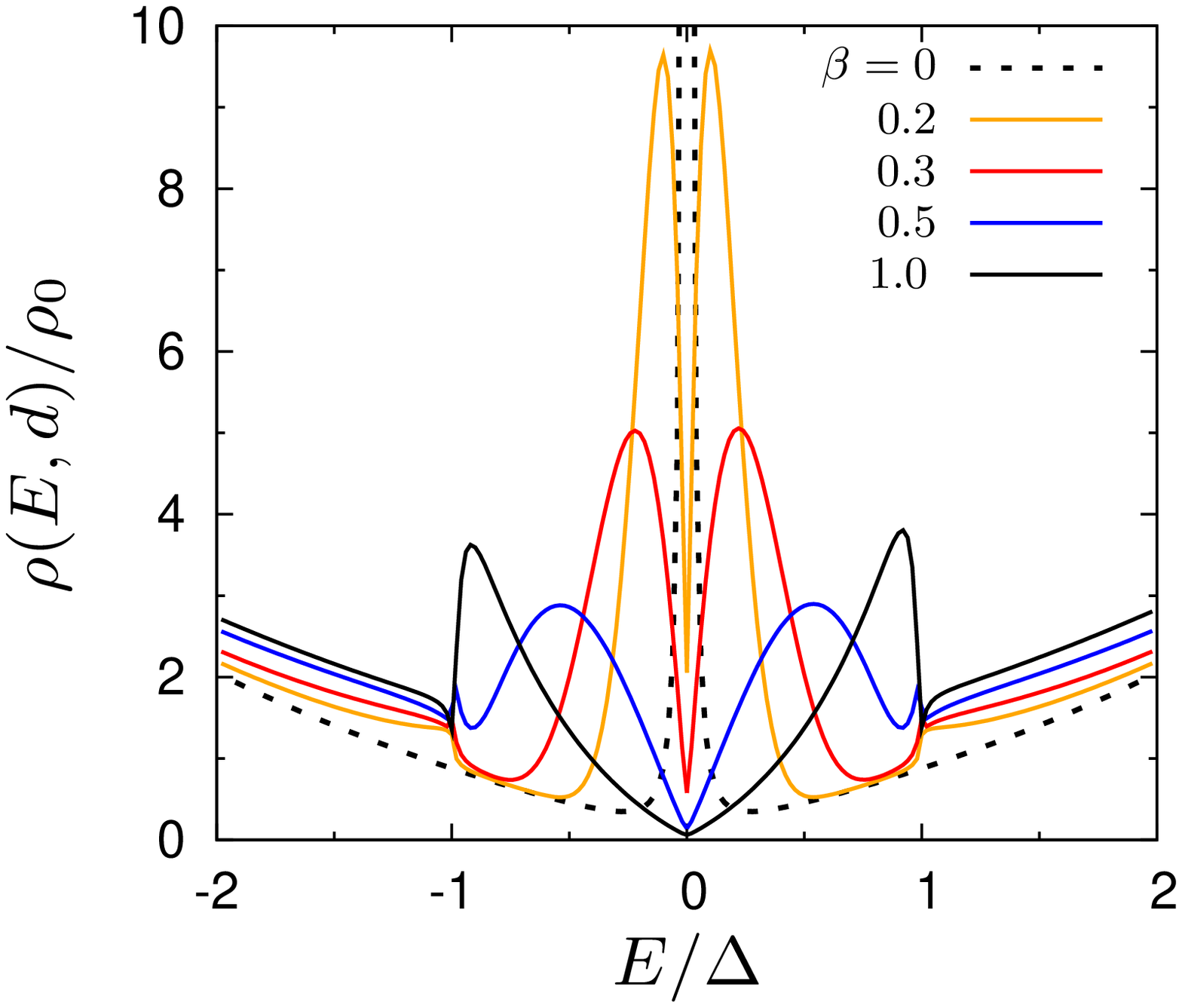}
\includegraphics[width=8.cm]{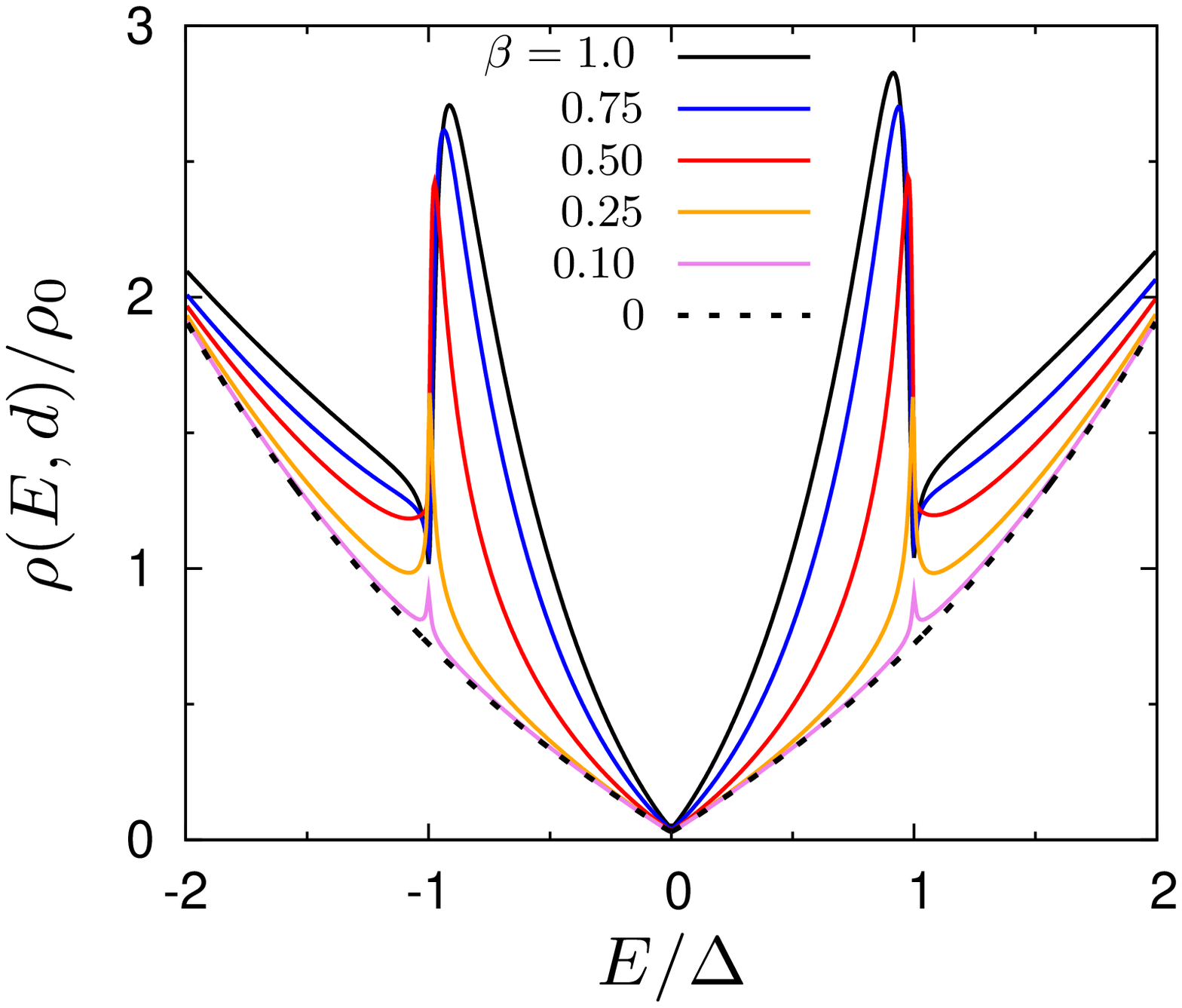}
\caption{(Color online) LDOS profiles, at a distance $d=-0.41\xi$ inside the normal region with $E_F=0$, for different transparencies of the zigzag (left panel) and armchair (right panel) interface. In the zigzag case, the edge state splits and evolves into a band inside the superconducting gap when the transparency $\beta$ goes from $0$ (uncoupled case represented by a dashed black line) to $1$ (perfect transparency, solid black line). For the armchair case the superconducting proximity effect is manifested by the appearance of sharp peaks in the LDOS for energies $E \sim \Delta$.}
\label{fig:LDOS_zz}
\end{figure*}

As a test of the method we analyze in this section the LDOS for a graphene-superconductor junction. The electronic LDOS per unit length for a distance $x=x'\equiv d$ is defined as
\begin{equation}
\rho(E,d) = \frac{-1}{\pi} \int_{-\infty}^{\infty} dq \mbox{Tr} \left[ \mbox{Im} \left\{ \check{G}_{L \vert ee} \left(q,E;d,d\right) \right\} \right]
\label{eq:LDOS}
\end{equation}

The left panel of Fig. \ref{fig:LDOS} shows the LDOS for an armchair edge close to the interface ($d=-0.1 \xi$) obtained using the boundary conditions of vanishing the wave function or its derivative. For comparison we also show in this figure the results obtained for the TB model calculations of Ref. \cite{burset08}. The results of Fig. \ref{fig:LDOS} are normalized to the density of a bulk graphene layer with zero doping at $E=\Delta$, $\rho_0 = 2 \Delta /\hbar^2 v^2$ (see Appendix \ref{appendix_B}). The discrepancies between the different matching conditions tend to disappear when we move away from the interface, as it is shown in the right panel of Fig. \ref{fig:LDOS}. This panel also illustrates the oscillatory behavior of the LDOS inside the normal region of graphene around the bulk value (indicated by the dashed line). The period of the oscillation is given roughly by $\hbar v/E$ and the amplitude decreases with the distance to the interface. When studying the spatial evolution of the LDOS of an armchair layer of graphene, it is well known that there is an oscillatory behavior at the atomic scale \cite{brey06,burset08}. This is a direct result of the valley mixing that happens at an armchair edge. For the sake of clarity, the evolution of the LDOS with the direction transversal to the interface shown in the right panel of Fig. \ref{fig:LDOS} has been calculated for values $\left\vert x \right\vert/a = 3n$, with $n$ a positive integer. Similar profiles are reached for different multiplicities of the coordinate $x$. The oscillatory behavior at the atomic scale is shown in the inset of this panel for the boundary conditions of vanishing the wave function.

On the other hand, the left panel of Fig. \ref{fig:LDOS_zz} shows the LDOS for a zigzag edge at a distance inside the normal region $d = -0.4 \xi$. The results correspond to different values of the parameter $\beta$ controlling the interface transparency. For the uncoupled case with $\beta=0$, a localized edge state appears at $E=E_F$ (dashed line in Fig. \ref{fig:LDOS_zz}). However, for a non-zero transparency the edge state splits and evolves into a band inside the superconducting gap. As it was demonstrated in Ref. \cite{burset09}, this band is formed by the IBSs which give rise to the peaks in the DOS for energies around $E=\pm \Delta$ when $\beta = 1$.

It is important to emphasize that although the surface state at $E = E_{F}$ for zigzag edges has been thoroughly studied, our results demonstrate that the superconducting proximity effect splits this state and produces a shift that depends of the transparency between the normal and superconducting regions.

The LDOS for the armchair case with the same set of parameters is illustrated in the right panel of Fig. \ref{fig:LDOS_zz}. As can be observed there is no localized state for the uncoupled case ($\beta = 0$) and when $\beta$ increases, two peaks arise around $E=\pm \Delta$ \cite{burset08}.

\subsection{Induced pairing correlations}

The coupling with the superconductor induces correlations between electron and hole-like excitations in the normal region by the proximity effect. We can map the spatial variation of these correlations using the Green function of the coupled system. The element $\check{G}_{L \vert eh,AA} (x,x',y)$ corresponds to the injection of electron excitations at the point $x'$ on the $y=0$ axis and its propagation as a hole excitation to the rest of the plane ($x,y$) after an electron-hole conversion at the interface. The probability $P_{e\rightarrow h}$ that an electron injected at $(x',0)$ would be converted into a hole at $(x,y)$ is proportional to
\begin{equation}
P_{e\rightarrow h} \propto \left\vert \int_{-\infty}^{\infty} dq e^{iqy} \check{G}_{L \vert eh,S S'} \left(q,E;x,x' \right) \right\vert^2 ,
\label{eq:spatial}
\end{equation}
with $S,S'=A,B$. 

\begin{figure}
\includegraphics[width=7.8cm]{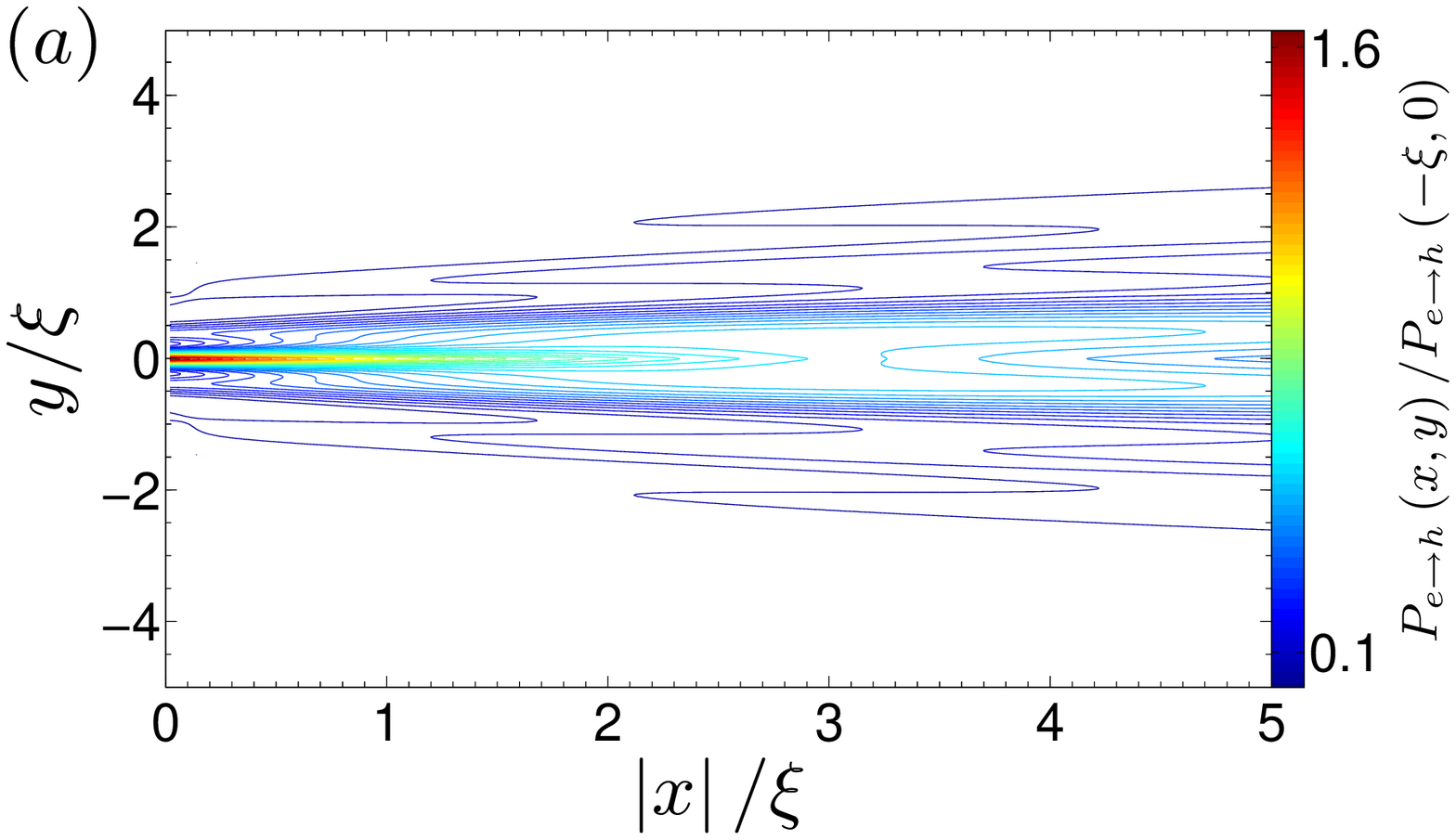}
\includegraphics[width=7.8cm]{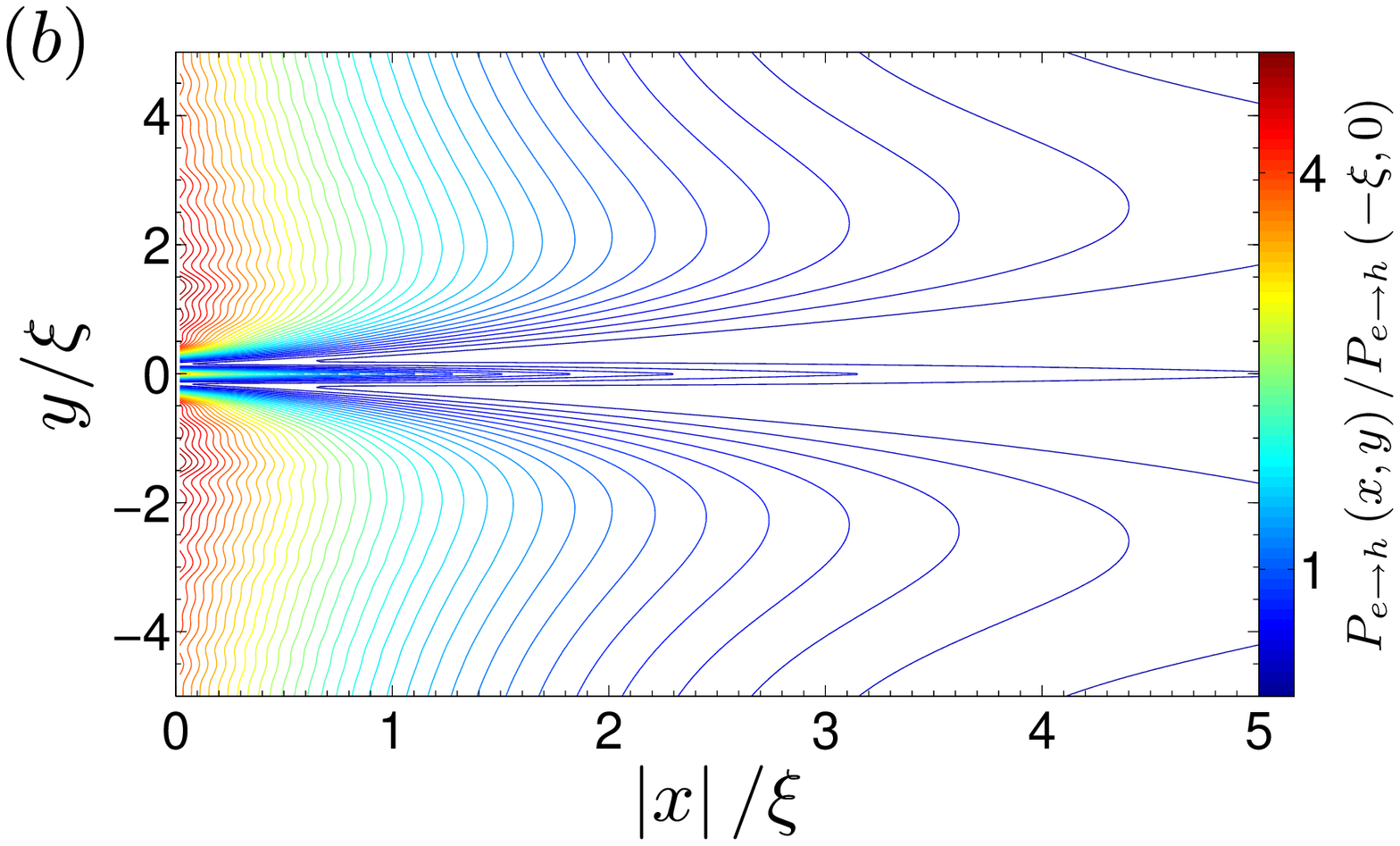}
\includegraphics[width=7.8cm]{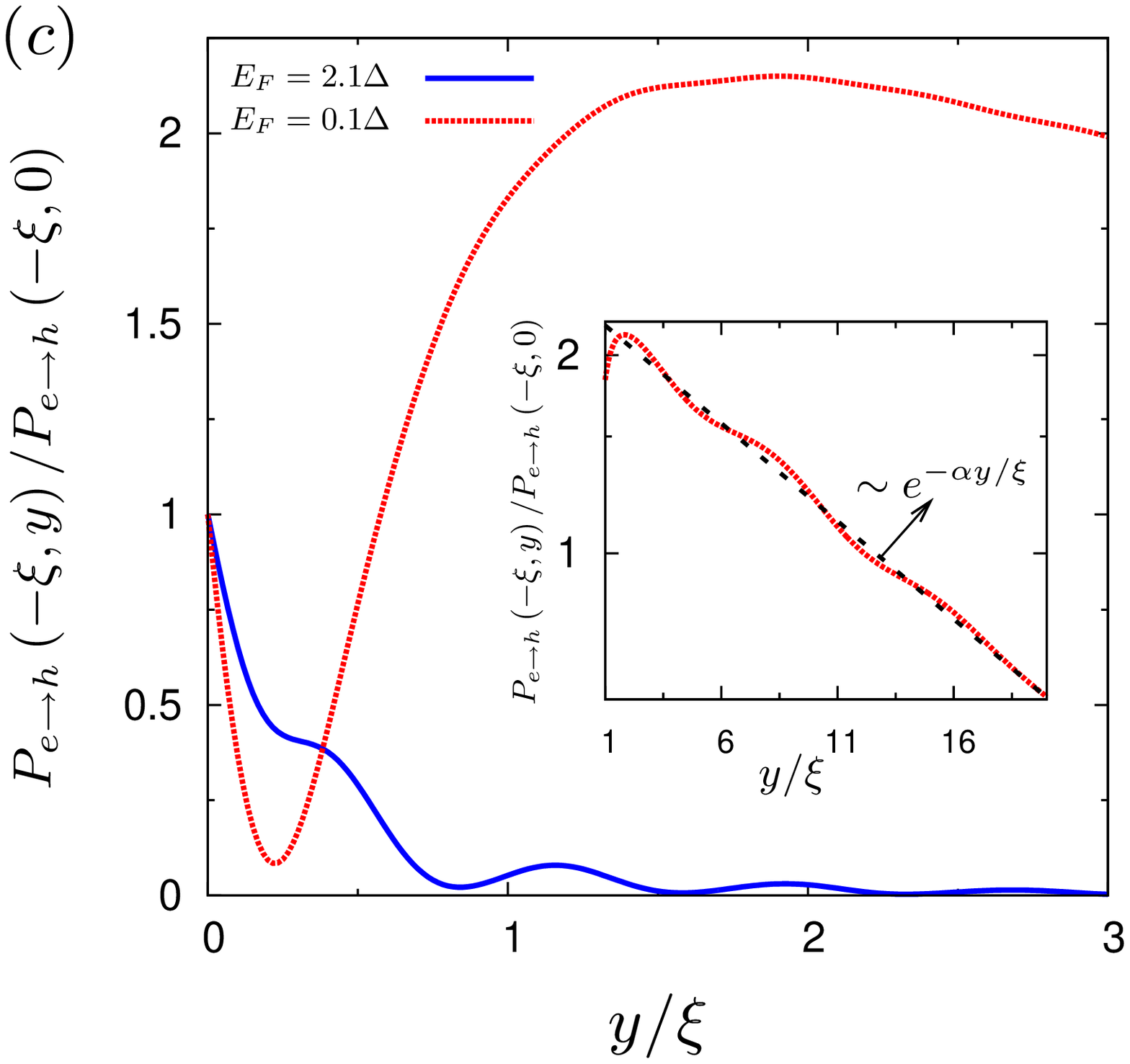}
\caption{(Color online) Spatial mapping of induced pairing correlations in the normal region of graphene. The results show the squared absolute value of the Fourier transform of $\check{G}_{L \vert eh,AA}$ for an incident energy $E=0.75\Delta$ and $x'=-\xi$. The panel on the left shows the case $E_F > \Delta$ with $E_F = 2.1\Delta$, in which retro-reflection is enhanced. The right panel shows the opposite case $E_F < \Delta$, with $E_F=0.1\Delta$, in which specular reflection dominates.}
\label{fig:spatial}
\end{figure}

In our study of the spatial variation of the superconducting correlations we set $x'=-\xi$. We find it interesting to explore the change in the induced pair correlations for electrons injected within the gap ($E<\Delta$) when the doping level varies from $E_F \gg \Delta$ (left panel of Fig. \ref{fig:spatial}) to $E_F \ll \Delta$ (right panel). As it was demonstrated in Ref. \cite{beenakker06}, these two limiting cases correspond to the regimes in which Andreev reflection exhibits respectively a retro or a specular character. In these limits and in the case of perfect transparency, the main difference between an armchair edge and a zigzag one is that the atomic oscillations happen along the $x$ or $y$ axis, respectively. Thus, we only analyze the results for an armchair edge with a fixed multiplicity. In Fig. \ref{fig:spatial} these two cases are shown for an incident energy $E=0.75 \Delta$. In the case $E_F > \Delta$, shown in the left panel, electron-hole conversion occurs mainly for $y \simeq 0$ decaying fast for $y \gtrsim \xi$. This confinement of the pairing correlations on the $y$ direction is a consequence of the underlying diffraction pattern for the injected electrons at one point which has a spatial extension of the order of $\hbar v/E_F$. The behavior is drastically different in the limit $E_F<E<\Delta$ for which specular reflection is enhanced (right panel). In addition to the electron-hole conversion for normal incidence, the results exhibit the presence of long-range superconducting correlations along the $y$ axis, provided that $x \sim \xi$. These long-range correlations can be directly associated with the presence of IBSs which in this regime have a spatial extension inside the normal region which is much larger than $\xi$. At the same time, the extended correlations can be interpreted as a signature of divergent Andreev reflection trajectories characteristic of the regime $E_F \ll \Delta$.

\section{Conclusions}
In this paper we have developed a method that allows to obtain the Green function in inhomogeneous graphene systems with well defined edges, such as nanoribbons and graphene-superconductor interfaces. In this approach the asymptotic solutions of the Bogoliubov-de Gennes-Dirac equations satisfying the appropriate boundary conditions are used to analytically build the Green functions associated to the quasiparticles in the system. 

Within this formalism we have analyzed the LDOS of a normal-superconducting graphene junction and showed how it is affected by the type of edge and the interface transparency. In particular, for the zigzag case, when the parameter $\beta$ (which controls the transparency of the junction) increases, the localized edge state appearing at $E=E_F$ splits and evolves into symmetric bands within the superconducting gap. These bands correspond to the interface bound states already described in a previous work \cite{burset09}.

Furthermore, we have studied the spatial distribution of the induced pairing correlations inside the normal region. This correlations illustrate the crossover between retro and specular Andreev reflection. We indicate the appearance of long-range correlations between distant points on the graphene layer close to the interface due to the presence of IBSs. These superconducting correlations are enhanced in the regime corresponding to specular Andreev reflection.

We would like to conclude stressing that the analytical results obtained in the present work could be very useful for the study of transport properties in more complex hybrid systems which combine finite normal and superconducting graphene regions. Work along these lines is under progress.

\acknowledgments
Financial support from Spanish MICINN under
contracts FIS2005-06255 and FIS2008-04209 and by DIB from Universidad
Nacional de Colombia is acknowledged.

\appendix

\section{Green's functions for zigzag edges. One valley description.}
\label{appendix_A}

For the case of a graphene sheet with zigzag edges, we have set the Dirac points to be at $\mathbf{K}_{\pm}=\left(0,\pm K \right)$. Thus, the Hamiltonian on each valley is given by $\hat{H}_{\pm}^{\text{zz}} = \hbar v \left[ k \hat{\sigma}_x \pm q \hat{\sigma}_y \right] -E_F \hat{\sigma}_0$. The Hamiltonian of the full system is related with the one given in Eq. (\ref{eq:hamiltonian}) by $\check{H}_{\text{zz}}=\check{T}
\check{H}\check{T}$, with 
\begin{equation}
\check{T}=\check{T}^{-1}=\left( 
\begin{array}{cc}
1 & 0 \\ 
0 & \hat{\sigma}_{z}
\end{array}
\right) .
\end{equation}
The matrix $\check{\gamma}=\hat{\tau}_{x}\otimes \hat{\sigma}_{x}$ is then transformed as $\check{\gamma}_{\text{zz}}=\check{T}
\check{\gamma}\check{T}=\hat{\tau}_y \otimes \hat{\sigma}_y$.
When we apply this matrix to $\check{H}_{\text{zz}}$ we have 
$\check{\gamma}_{\text{zz}} \left( \check{H}_{\text{zz}}\left( \mathbf{k}\right) \right) \check{\gamma}_{\text{zz}}^{-1} = \check{H}_{\text{zz}}(-\mathbf{k})$, which corresponds to a parity transformation. 

Since the zigzag boundary conditions do not mix valleys we can work only with one valley ($\mathbf{K}_+$) using Hamiltonian $\hat{H}_{+}^{\text{zz}} = \hat{H}_{+}$. For this Hamiltonian the effect of the parity transformation, restricted to the sublattice subspace, is equivalent to set $\hat{\gamma}=\hat{\sigma}_z$.

\section{Bulk Green's functions.}
\label{appendix_B}

If we set all the reflection amplitudes to zero in Eq. \eqref{eq:FdG_arm_f} we obtain the bulk solution for an infinite graphene layer
\begin{equation}
\begin{array}{c}
\hat{G}^{\text{bulk}}_{\phi} (x,x^{\prime }) = 
\frac{-i}{2\hbar v\cos{\alpha}} \\
\times \left\{ 
\begin{array}{cr}
e^{i\left( K+k\right) \left\vert x-x^{\prime }\right\vert }\varphi
_{1}\varphi _{1}^{\dagger }+e^{-i\left( K-k\right) \left\vert x-x^{\prime
}\right\vert }\varphi _{2}\varphi _{2}^{\dagger } & ;x<x^{\prime } \\ 
e^{i\left( K+k\right) \left\vert x-x^{\prime }\right\vert }\varphi
_{2}\varphi _{2}^{\dagger }+e^{-i\left( K-k\right) \left\vert x-x^{\prime
}\right\vert }\varphi _{1}\varphi _{1}^{\dagger } & ;x>x^{\prime }
\end{array}
\right.
\end{array}
\label{eq:FdG_bulk}
\end{equation}
Thus, the local propagator for a bulk of graphene is
\begin{equation}
\begin{array}{c}
\hat{G}^{\text{bulk}}_{\phi} (x=x^{\prime }) = \frac{-i}{2\hbar v\cos{\alpha}} \left[ \varphi _{1}\varphi _{1}^{\dagger } + \varphi _{2}\varphi _{2}^{\dagger } \right] = \\
\frac{-i}{\hbar v} \left[ \cos^{-1}{\alpha} \hat{\sigma}_0 + \tan{\alpha} \hat{\sigma}_y \right].
\end{array}
\end{equation}
Furthermore, the bulk LDOS is obtained integrating this simple result, thus giving $\rho_{\text{bulk}}(E) = 2\left(E+E_F \right)/\hbar^2 v^2$.

On the other hand, vanishing the reflection coefficient ($r$) in Eq. (\ref{eq:FdG_arm_super}) we reach the bulk solution for an infinite superconducting region. The local valley superposed Green function is then written as
\begin{equation}
\begin{aligned}
\check{G}&^{S,\text{bulk}}_{\phi}(x=x^{\prime }) = \frac{-i}{2\hbar v} \left\{ 
\left( \cos^{-1}{\alpha^S_e} \hat{\sigma}_0 + \tan{\alpha^S_e} \hat{\sigma}_y \right) \right. \\
& \left. \otimes \frac{1}{\Omega } \left( E \hat{\tau}_{0} + \Delta \hat{\tau}_{y} + \hat{\tau}_{z} \right) \right. \\
& \left. + \left( \cos^{-1}{\alpha^S_h} \hat{\sigma}_0 + \tan{\alpha^S_h} \hat{\sigma}_y \right) \otimes \frac{1}{\Omega } \left( E \hat{\tau}_{0} + \Delta \hat{\tau}_{y} - \hat{\tau}_{z} \right) \right\}.
\end{aligned}
\label{eq:FdG_bulk_super}
\end{equation}

In the heavily doped limit, the Green's function loses all structure in sublattice space,
\begin{equation}
\check{G}^{S,\text{bulk}}_{\phi}(x=x^{\prime }) \approx \frac{-i}{\hbar v \Omega}
\left( E \hat{\tau}_{0} + \Delta \hat{\tau}_{y} \right)
 ,
\end{equation}
and is equivalent to the bulk BCS Green function.


\begin{thebibliography}{99}
\bibitem{tkachov09} A notably exception is the work by G. Tkachov, Phys. Rev. B {\bf 79}, 045429 (2009), where the Green functions for normal graphene layers with zigzag edges have been derived.
\end{thebibliography}
\end{document}